\documentclass[12pt]{article}
\usepackage{amssymb}
\usepackage{amsmath}
\usepackage{amsfonts}

\begin{document}

\author{C. Bizdadea\thanks{%
e-mail address: bizdadea@central.ucv.ro}, C. C. Ciob\^{\i}rc\u{a}\thanks{%
e-mail address: ciobarca@central.ucv.ro}, I. Negru\thanks{%
e-mail address: inegru@central.ucv.ro}, S. O. Saliu\thanks{%
e-mail address: osaliu@central.ucv.ro} \\
Faculty of Physics, University of Craiova\\
13 A. I. Cuza Str., Craiova 200585, Romania}
\title{Couplings between a single massless tensor field with
the mixed symmetry (3,1) and one vector field}
\maketitle

\begin{abstract}
Under the hypotheses of smoothness in the coupling constant,
locality, Lorentz covariance, and Poincar\'{e} invariance of the
deformations, combined with the preservation of the number of
derivatives on each field, the consistent interactions between a
single free massless tensor gauge field with the mixed symmetry of a
two-column Young diagram of the type (3,1) and one Abelian vector
field have been investigated. The computations are done with the
help of the deformation theory based on a cohomological approach, in
the context of the antifield-BRST formalism. The main result is that
there exist nontrivial cross-couplings between these types of fields
in five spatiotemporal dimensions, which break the PT invariance and
allow for the deformation of the gauge transformations of the vector
field, but not of the gauge algebra.

\end{abstract}

\section{Introduction}

Tensor fields in ``exotic" representations
of the Lorentz group, characterized by a mixed Young symmetry type~\cite%
{curt1,curt2,aul,labast1,labast2,burd,zinov1}, held the attention lately on
some important issues, like the dual formulation of field theories of spin
two or higher~\cite%
{dualsp1,dualsp2,dualsp2a,dualsp2b,dualsp3,dualsp4,dualsp5}, the
impossibility of consistent cross-interactions in the dual formulation of
linearized gravity~\cite{lingr}, or a Lagrangian first-order approach~\cite%
{zinov2,zinov3} to some classes of massless or partially massive mixed
symmetry type tensor gauge fields, suggestively resembling to the tetrad
formalism of General Relativity. An important matter related to mixed
symmetry type tensor fields is the study of their consistent interactions,
among themselves as well as with higher-spin gauge theories~\cite%
{high1,high2,high3,high4,7,9,kk,3,4}. The most efficient approach to
this problem is the cohomological one, based on the deformation of
the solution to the master equation~\cite{def}. The purpose of this
paper is to investigate the consistent interactions between a single
free massless tensor gauge field $t_{\lambda \mu \nu \vert \alpha }$
with the mixed symmetry of a two-column Young diagram of the type
$(3,1)$ and one Abelian vector field $A_{\mu }$. It is worth
mentioning the duality of the free massless tensor gauge field
$t_{\lambda \mu \nu \vert \alpha }$ to the Pauli-Fierz theory in
$D=6$ dimensions and, in this respect, the recent developments
concerning the dual formulations of linearized gravity from the
perspective of $M$-theory~\cite{mth1,mth2,mth3}. Our analysis relies
on the deformation of the solution to the master equation by means
of cohomological techniques with the help of the local BRST
cohomology, whose component in the $(3,1)$ sector has been reported
in detail in~\cite{noijhep31}. Under the hypotheses of smoothness in
the coupling constant, locality, Lorentz covariance, and
Poincar\'{e} invariance of the deformations, combined with the
preservation of the number of derivatives on each field, we prove
that there exists a case where the deformation of the solution to
the master equation provides nontrivial cross-couplings. This case
corresponds to a five-dimensional space-time and is described by a
deformed solution that stops at order two in the coupling constant.
The interacting Lagrangian action contains only mixing-component
terms of order one and two in the coupling constant. At the level of
the gauge transformations, only those of the vector fields are
modified at order one in the coupling constant with a term linear in
the antisymmetrized first-order derivatives of some gauge parameters
from the $(3,1)$ sector such that the gauge algebra and the
reducibility structure of the coupled model are not modified during
the deformation procedure, being the same like in the case of the
starting free action. It is interesting to note that if we require
the PT invariance of the deformed theory, then no interactions
occur. Although it is not possible to construct interactions that
deform the gauge algebra, our result is interesting since this seems
to be the first case where mixed symmetry type tensor fields allow
nontrivial cross-couplings.

\section{Free model. BRST symmetry}

We begin with the Lagrangian action
\begin{eqnarray}
&&S_{0}\left[ t_{\lambda \mu \nu \vert \alpha },A_{\mu }\right] =\int
d^{D}x\left\{ \frac{1}{2}\left[ \left( \partial ^{\rho }t^{\lambda \mu \nu
\vert \alpha }\right) \left( \partial _{\rho }t_{\lambda \mu \nu \vert
\alpha }\right) -\left( \partial _{\alpha }t^{\lambda \mu \nu \vert \alpha
}\right) \left( \partial ^{\beta }t_{\lambda \mu \nu \vert \beta }\right) %
\right] \right.  \notag \\
&&-\frac{3}{2}\left[ \left( \partial _{\lambda }t^{\lambda \mu \nu \vert
\alpha }\right) \left( \partial ^{\rho }t_{\rho \mu \nu \vert \alpha
}\right) +\left( \partial ^{\rho }t^{\lambda \mu }\right) \left( \partial
_{\rho }t_{\lambda \mu }\right) \right] +3\left( \partial _{\alpha
}t^{\lambda \mu \nu \vert \alpha }\right) \left( \partial _{\lambda }t_{\mu
\nu }\right)  \notag \\
&&\left. +3\left( \partial _{\rho }t^{\rho \mu }\right) \left( \partial
^{\lambda }t_{\lambda \mu }\right) -\frac{1}{4}F_{\mu \nu }F^{\mu \nu
}\right\} \equiv S_{0}^{\mathrm{t}}\left[ t_{\lambda \mu \nu \vert \alpha }%
\right] +S_{0}^{\mathrm{A}}\left[ A_{\mu }\right] ,  \label{tv1}
\end{eqnarray}%
in $D\geq 5$ spatiotemporal dimensions. The massless tensor field $%
t_{\lambda \mu \nu \vert \alpha }$ has the mixed symmetry $(3,1)$ and hence
transforms according to an irreducible representation of $GL(D,\mathbb{R})$
corresponding to a 4-cell Young diagram with two columns and three rows. It
is thus completely antisymmetric in its first three indices and satisfies
the identity $t_{[\lambda \mu \nu \vert \alpha ]}\equiv 0$. The field
strength of the vector field $A_{\mu }$ is defined in the standard manner by
\begin{equation}
F_{\mu \nu }=\partial _{\mu }A_{\nu }-\partial _{\nu }A_{\mu }\equiv
\partial _{\left[ \mu \right. }A_{\left. \nu \right] }.  \label{abfstr}
\end{equation}%
Everywhere in this paper it is understood that the notation $[\lambda \cdots
\alpha ]$ signifies complete antisymmetry with respect to the (Lorentz)
indices between brackets, with the conventions that the minimum number of
terms is always used and the result is never divided by the number of terms.
The trace of $t_{\lambda \mu \nu \vert \alpha }$ is defined by $t_{\lambda
\mu }=\sigma ^{\nu \alpha }t_{\lambda \mu \nu \vert \alpha }$ and it is
obviously an antisymmetric tensor. Everywhere in this paper we employ the
flat Minkowski metric of `mostly plus' signature $\sigma ^{\mu \nu }=\sigma
_{\mu \nu }=(-,++++\cdots )$.

A generating set of gauge transformations for the action (\ref{tv1}) can be
taken of the form
\begin{eqnarray}
\delta _{\epsilon ,\chi }t_{\lambda \mu \nu \vert \alpha } &=&-3\partial _{
\left[ \lambda \right. }\epsilon _{\left. \mu \nu \alpha \right] }+4\partial
_{\left[ \lambda \right. }\epsilon _{\left. \mu \nu \right] \alpha
}+\partial _{\left[ \lambda \right. }\chi _{\left. \mu \nu \right] \vert
\alpha },  \label{tv7a} \\
\delta _{\epsilon }A_{\mu } &=&\partial _{\mu }\epsilon ,  \label{tv7b}
\end{eqnarray}%
where the gauge parameters $\epsilon _{\lambda \mu \nu }$ determine a
completely antisymmetric tensor, the other set of gauge parameters displays
the mixed symmetry $\left( 2,1\right) $, such that they are antisymmetric in
the first two indices and satisfy the identity $\chi _{\left[ \mu \nu \vert
\alpha \right] }\equiv 0$, and the gauge parameter $\epsilon $ is a scalar.
The generating set of gauge transformations (\ref{tv7a})--(\ref{tv7b}) is
off-shell, second-stage reducible, the accompanying gauge algebra being
obviously Abelian. More precisely, the gauge transformations (\ref{tv7a})
are off-shell, second-stage reducible. This is because: 1. 1. If in (\ref%
{tv7a}) we make the transformations
\begin{eqnarray}
\epsilon _{\mu \nu \alpha } &\rightarrow &\epsilon _{\mu \nu \alpha
}^{\left( \omega ,\psi \right) }=-\frac{1}{2}\partial _{\left[ \mu \right.
}\omega _{\left. \nu \alpha \right] },  \label{tv12} \\
\chi _{\mu \nu \vert \alpha } &\rightarrow &\chi _{\mu \nu \vert \alpha
}^{\left( \omega ,\psi \right) }=\partial _{\left[ \mu \right. }\psi
_{\left. \nu \right] \alpha }+2\partial _{\alpha }\omega _{\mu \nu
}-\partial _{\left[ \mu \right. }\omega _{\left. \nu \right] \alpha },
\label{tv12a}
\end{eqnarray}%
with $\omega _{\nu \alpha }$ antisymmetric and $\psi _{\nu \alpha }$
symmetric (but otherwise arbitrary), then the gauge variation of the tensor
field identically vanishes $\delta _{\epsilon ^{\left( \omega ,\psi \right)
},\chi ^{\left( \omega ,\psi \right) }}t_{\lambda \mu \nu \vert \alpha
}\equiv 0$. 2. If in (\ref{tv12})--(\ref{tv12a}) we perform the changes
\begin{eqnarray}
\omega _{\nu \alpha } &\rightarrow &\omega _{\nu \alpha }^{\left( \theta
\right) }=\partial _{\left[ \mu \right. }\theta _{\left. \nu \right] },
\label{tv15} \\
\psi _{\nu \alpha } &\rightarrow &\psi _{\nu \alpha }^{\left( \theta \right)
}=-3\partial _{\left( \mu \right. }\theta _{\left. \nu \right) },
\label{tv15a}
\end{eqnarray}%
with $\theta _{\nu }$ an arbitrary vector field, where $\left( \mu \nu
\cdots \right) $ signifies symmetrization with respect to the indices
between parentheses without normalization factors, then the transformed
gauge parameters (\ref{tv12})--(\ref{tv12a}) identically vanish $\epsilon
_{\mu \nu \alpha }^{\left( \omega ^{\left( \theta \right) },\psi ^{\left(
\theta \right) }\right) }\equiv 0$, $\chi _{\mu \nu \vert \alpha }^{\left(
\omega ^{\left( \theta \right) },\psi ^{\left( \theta \right) }\right)
}\equiv 0$. 3. There is no non-vanishing local transformation of $\theta
_{\nu }$ that simultaneously annihilates $\omega _{\nu \alpha }^{\left(
\theta \right) }$ and $\psi _{\nu \alpha }^{\left( \theta \right) }$ of the
form (\ref{tv15})--(\ref{tv15a}) and hence no further local reducibility
identity.

The field equations associated with action (\ref{tv1}) are
\begin{equation}
\Sigma :\frac{\delta S_{0}}{\delta t_{\lambda \mu \nu \vert \alpha }}\equiv
-T^{\lambda \mu \nu \vert \alpha }\approx 0,\quad \frac{\delta S_{0}}{\delta
A_{\mu }}\equiv \partial _{\nu }F^{\nu \mu }\approx 0,  \label{tv16}
\end{equation}%
where (minus) the Euler-Lagrange derivatives $T^{\lambda \mu \nu \vert
\alpha }$ have the form
\begin{eqnarray}
T^{\lambda \mu \nu \vert \alpha } &=&\Box t^{\lambda \mu \nu \vert \alpha
}-\partial _{\rho }\left( \partial ^{\left[ \lambda \right. }t^{\left. \mu
\nu \right] \rho \vert \alpha }+\partial ^{\alpha }t^{\lambda \mu \nu \vert
\rho }\right) +\partial ^{\alpha }\partial ^{\left[ \lambda \right.
}t^{\left. \mu \nu \right] }  \notag \\
&&+\sigma ^{\alpha \left[ \lambda \right. }\left( \partial _{\rho }\left(
\partial _{\beta }t^{\left. \mu \nu \right] \rho \vert \beta }-\partial
^{\mu }t^{\left. \nu \right] \rho }\right) -\Box t^{\left. \mu \nu \right]
}\right) .  \label{tv17}
\end{eqnarray}%
The notation $\approx $ means here the weak equality symbol. The tensor $%
T^{\lambda \mu \nu \vert \alpha }$ has the same properties like the field $%
t^{\lambda \mu \nu \vert \alpha }$, being antisymmetric in its first three
indices and satisfying the identity $T^{\left[ \lambda \mu \nu \vert \alpha %
\right] }\equiv 0$. Its trace
\begin{equation}
T^{\lambda \mu }\equiv \sigma _{\nu \alpha }T^{\lambda \mu \nu \vert \alpha
}=\left( 4-D\right) \left( \Box t^{\lambda \mu }+\partial _{\rho }\left(
\partial ^{\left[ \lambda \right. }t^{\left. \mu \right] \rho }-\partial
_{\alpha }t^{\lambda \mu \rho \vert \alpha }\right) \right) ,
\label{tv18aaaa}
\end{equation}%
is an antisymmetric tensor.

The most general quantities invariant under the gauge transformations (\ref%
{tv7a})--(\ref{tv7b}) are functions of the curvature tensor
\begin{equation}
K^{\lambda \mu \nu \xi \vert \alpha \beta }=\partial ^{\alpha }\partial ^{
\left[ \lambda \right. }t^{\left. \mu \nu \xi \right] \vert \beta }-\partial
^{\beta }\partial ^{\left[ \lambda \right. }t^{\left. \mu \nu \xi \right]
\vert \alpha },  \label{tv20}
\end{equation}%
of the field strength (\ref{abfstr}) as well as of their spatiotemporal
derivatives of all orders. The curvature tensor exhibits the mixed symmetry $%
\left( 4,2\right) $, such that it is separately antisymmetric in its first
four indices and respectively in the last two ones and fulfills the
(algebraic) Bianchi I identity $K^{\left[ \lambda \mu \nu \xi \vert \alpha %
\right] \beta }\equiv 0$. Meanwhile, the curvature tensor obeys two types of
(differential) Bianchi II identities
\begin{equation}
\partial ^{\left[ \kappa \right. }K^{\left. \lambda \mu \nu \xi \right]
\vert \alpha \beta }\equiv 0,\quad K^{\lambda \mu \nu \xi \vert \left[
\alpha \beta ,\gamma \right] }\equiv 0,  \label{bianchi2}
\end{equation}%
where we made the notation $K^{\lambda \mu \nu \xi \vert \alpha \beta
,\gamma }\equiv \partial ^{\gamma }K^{\lambda \mu \nu \xi \vert \alpha \beta
}$. Its traces are defined through
\begin{eqnarray}
K^{\lambda \mu \nu \vert \alpha } &\equiv &\sigma _{\xi \beta }K^{\lambda
\mu \nu \xi \vert \alpha \beta }=\Box t^{\lambda \mu \nu \vert \alpha
}-\partial _{\rho }\partial ^{\left[ \lambda \right. }t^{\left. \mu \nu %
\right] \rho \vert \alpha }  \notag \\
&&-\partial ^{\alpha }\left( \partial _{\beta }t^{\lambda \mu \nu \vert
\beta }-\partial ^{\left[ \lambda \right. }t^{\left. \mu \nu \right]
}\right) ,  \label{tv23}
\end{eqnarray}%
\begin{equation}
K^{\lambda \mu }\equiv \sigma _{\nu \alpha }K^{\lambda \mu \nu \vert \alpha
}=2\left( \Box t^{\lambda \mu }+\partial _{\rho }\left( \partial ^{\left[
\lambda \right. }t^{\left. \mu \right] \rho }-\partial _{\alpha }t^{\lambda
\mu \rho \vert \alpha }\right) \right) ,  \label{tv24}
\end{equation}%
with $K^{\lambda \mu \nu \vert \alpha }$ having the same mixed symmetry like
the original tensor field $t_{\lambda \mu \nu \vert \alpha }$ and $%
K^{\lambda \mu }$ being antisymmetric.

It is interesting to note that if $\bar{T}^{\lambda \mu \nu \vert \alpha }$
is a covariant tensor field with the mixed symmetry $\left( 3,1\right) $,
which simultaneously satisfies the equations
\begin{equation}
\partial _{\lambda }\bar{T}^{\lambda \mu \nu \vert \alpha }=0,\quad \partial
_{\alpha }\bar{T}^{\lambda \mu \nu \vert \alpha }=0,  \label{tv32c}
\end{equation}%
then there exists a tensor $\bar{\Phi}^{\rho \lambda \mu \nu \vert \beta
\alpha }$ with the mixed symmetry of the curvature tensor, such that
\begin{equation}
\bar{T}^{\lambda \mu \nu \vert \alpha }=\partial _{\xi }\partial _{\beta }%
\bar{\Phi}^{\lambda \mu \nu \xi \vert \alpha \beta }.  \label{tv32d}
\end{equation}%
(A constant solution $C^{\lambda \mu \nu \vert \alpha }$ is excluded
from covariance arguments due to the mixed symmetry $\left(
3,1\right) $.)

The construction of the antifield-BRST symmetry for this free theory debuts
with the identification of the algebra on which the BRST differential $s$
acts. The generators of the BRST algebra are of two kinds: fields/ghosts and
antifields. The ghost spectrum for the model under study comprises the
fermionic ghosts $\left\{ \eta _{\lambda \mu \nu },\mathcal{G}_{\mu \nu
\vert \alpha },\eta \right\} $ associated with the gauge parameters $\left\{
\epsilon _{\lambda \mu \nu },\chi _{\mu \nu \vert \alpha },\epsilon \right\}
$ from (\ref{tv7a})--(\ref{tv7b}), the bosonic ghosts for ghosts $\left\{
C_{\mu \nu },\mathcal{C}_{\nu \alpha }\right\} $ due to the first-stage
reducibility parameters $\left\{ \omega _{\mu \nu },\psi _{\nu \alpha
}\right\} $ in (\ref{tv12})--(\ref{tv12a}), and also the fermionic ghost for
ghost for ghost $C_{\nu }$ corresponding to the second-stage reducibility
parameter $\theta _{\nu }$ in (\ref{tv15})--(\ref{tv15a}). In order to make
compatible the behavior of the gauge and reducibility parameters with that
of the accompanying ghosts, we ask that $\eta _{\lambda \mu \nu }$ and $%
C_{\mu \nu }$ are completely antisymmetric, $\mathcal{G}_{\mu \nu \vert
\alpha }$ obeys the analogue of the properties fulfilled by the gauge
parameters $\chi _{\mu \nu \vert \alpha }$, and $\mathcal{C}_{\nu \alpha }$
is symmetric. The antifield spectrum is organized into the antifields $%
\left\{ t^{\ast \lambda \mu \nu \vert \alpha },A^{\ast \mu }\right\} $ of
the original tensor fields, together with those of the ghosts, $\left\{ \eta
^{\ast \lambda \mu \nu },\mathcal{G}^{\ast \mu \nu \vert \alpha },\eta
^{\ast }\right\} $, $\left\{ C^{\ast \mu \nu },\mathcal{C}^{\ast \nu \alpha
}\right\} $, and respectively $C^{\ast \nu }$, of statistics opposite to
that of the associated fields/ghosts. It is understood that $t^{\ast \lambda
\mu \nu \vert \alpha }$ exhibits the same mixed-symmetry properties like $%
t_{\lambda \mu \nu \vert \alpha }$ and similarly with respect to $\eta
^{\ast \lambda \mu \nu }$, $\mathcal{G}^{\ast \mu \nu \vert \alpha }$, $%
C^{\ast \mu \nu }$, and $\mathcal{C}^{\ast \nu \alpha }$. For subsequent
purpose, we denote the trace of $t^{\ast \lambda \mu \nu \vert \alpha }$ by $%
t^{\ast \lambda \mu }$, being understood that it is antisymmetric.

Since both the gauge generators and reducibility functions for this model
are field-independent, it follows that the BRST differential $s$ simply
reduces to
\begin{equation}
s=\delta +\gamma ,  \label{tv39}
\end{equation}%
where $\delta $ represents the Koszul-Tate differential, graded by the
antighost number $\mathrm{agh}$ ($\mathrm{agh}\left( \delta \right) =-1$)
and $\gamma $ stands for the exterior derivative along the gauge orbits,
whose degree is named pure ghost number $\mathrm{pgh}$ ($\mathrm{pgh}\left(
\gamma \right) =1$). These two degrees do not interfere ($\mathrm{agh}\left(
\gamma \right) =0$, $\mathrm{pgh}\left( \delta \right) =0$). The overall
degree that grades the BRST complex is known as the ghost number ($\mathrm{gh%
}$) and is defined like the difference between the pure ghost number and the
antighost number, such that $\mathrm{gh}\left( s\right) =\mathrm{gh}\left(
\delta \right) =\mathrm{gh}\left( \gamma \right) =1$. According to the
standard rules of the BRST method, the corresponding degrees of the
generators from the BRST complex are valued like
\begin{equation*}
\mathrm{pgh}\left( t_{\lambda \mu \nu \vert \alpha }\right) =0=\mathrm{pgh}%
\left( A_{\mu }\right) ,\quad \mathrm{pgh}\left( \eta _{\lambda \mu \nu
}\right) =\mathrm{pgh}\left( \mathcal{G}_{\mu \nu \vert \alpha }\right) =%
\mathrm{pgh}\left( \eta \right) =1,
\end{equation*}%
\begin{equation*}
\mathrm{pgh}\left( C_{\mu \nu }\right) =2=\mathrm{pgh}\left( \mathcal{C}%
_{\nu \alpha }\right) ,\quad \mathrm{pgh}\left( C_{\nu }\right) =3,\quad
\mathrm{pgh}\left( t^{\ast \lambda \mu \nu \vert \alpha }\right) =\mathrm{pgh%
}\left( A^{\ast \mu }\right) =0,
\end{equation*}%
\begin{equation*}
\mathrm{pgh}\left( \eta ^{\ast \lambda \mu \nu }\right) =\mathrm{pgh}\left(
\mathcal{G}^{\ast \mu \nu \vert \alpha }\right) =\mathrm{pgh}\left( \eta
^{\ast }\right) =0,
\end{equation*}%
\begin{equation*}
\mathrm{pgh}\left( C^{\ast \mu \nu }\right) =\mathrm{pgh}\left( \mathcal{C}%
^{\ast \nu \alpha }\right) =\mathrm{pgh}\left( C^{\ast \nu }\right) =0,
\end{equation*}%
\begin{equation*}
\mathrm{agh}\left( t_{\lambda \mu \nu \vert \alpha }\right) =\mathrm{agh}%
\left( A_{\mu }\right) =0,\quad \mathrm{agh}\left( \eta _{\lambda \mu \nu
}\right) =\mathrm{agh}\left( \mathcal{G}_{\mu \nu \vert \alpha }\right) =%
\mathrm{agh}\left( \eta \right) =0,
\end{equation*}%
\begin{equation*}
\mathrm{agh}\left( C_{\mu \nu }\right) =\mathrm{agh}\left( \mathcal{C}_{\nu
\alpha }\right) =\mathrm{agh}\left( C_{\nu }\right) =0,\quad \mathrm{agh}%
\left( t^{\ast \lambda \mu \nu \vert \alpha }\right) =1=\mathrm{agh}\left(
A^{\ast \mu }\right) ,
\end{equation*}%
\begin{equation*}
\mathrm{agh}\left( \eta ^{\ast \lambda \mu \nu }\right) =\mathrm{agh}\left(
\mathcal{G}^{\ast \mu \nu \vert \alpha }\right) =\mathrm{agh}\left( \eta
^{\ast }\right) =2,
\end{equation*}%
\begin{equation*}
\mathrm{agh}\left( C^{\ast \mu \nu }\right) =3=\mathrm{agh}\left( \mathcal{C}%
^{\ast \nu \alpha }\right) ,\quad \mathrm{agh}\left( C^{\ast \nu }\right) =4.
\end{equation*}%
Actually, (\ref{tv39}) is a decomposition of the BRST differential according
to the antighost number and it shows that $s$ contains only components of
antighost number equal to minus one and zero. The Koszul-Tate differential
is imposed to realize a homological resolution of the algebra of smooth
functions defined on the stationary surface of field equations (\ref{tv16})
and the exterior longitudinal derivative is related to the gauge symmetries (%
\ref{tv7a})--(\ref{tv7b}) of the action (\ref{tv1}) through its cohomology
at pure ghost number zero computed in the cohomology of $\delta $, which is
required to be the algebra of physical observables for the free model under
consideration. The actions of $\delta $ and $\gamma $ on the generators from
the BRST complex, which enforce all the above mentioned properties, are
given by
\begin{equation}
\gamma t_{\lambda \mu \nu \vert \alpha }=-3\partial _{\left[ \lambda \right.
}\eta _{\left. \mu \nu \alpha \right] }+4\partial _{\left[ \lambda \right.
}\eta _{\left. \mu \nu \right] \alpha }+\partial _{\left[ \lambda \right. }%
\mathcal{G}_{\left. \mu \nu \right] \vert \alpha },\quad \gamma A_{\mu
}=\partial _{\mu }\eta ,  \label{tv49}
\end{equation}%
\begin{equation}
\gamma \eta _{\lambda \mu \nu }=-\frac{1}{2}\partial _{\left[ \lambda
\right. }C_{\left. \mu \nu \right] },\quad \gamma \eta =0,  \label{tv50}
\end{equation}%
\begin{equation}
\gamma \mathcal{G}_{\mu \nu \vert \alpha }=2\partial _{\left[ \mu \right.
}C_{\left. \nu \alpha \right] }-3\partial _{\left[ \mu \right. }C_{\left.
\nu \right] \alpha }+\partial _{\left[ \mu \right. }\mathcal{C}_{\left. \nu %
\right] \alpha },  \label{tv51}
\end{equation}%
\begin{equation}
\gamma C_{\mu \nu }=\partial _{\left[ \mu \right. }C_{\left. \nu \right]
},\quad \gamma \mathcal{C}_{\nu \alpha }=-3\partial _{\left( \nu \right.
}C_{\left. \alpha \right) },\quad \gamma C_{\nu }=0,  \label{tv52}
\end{equation}%
\begin{equation}
\gamma t^{\ast \lambda \mu \nu \vert \alpha }=\gamma A^{\ast \mu }=\gamma
\eta ^{\ast \lambda \mu \nu }=\gamma \mathcal{G}^{\ast \mu \nu \vert \alpha
}=\gamma \eta ^{\ast }=0,  \label{tv53}
\end{equation}%
\begin{equation}
\gamma C^{\ast \mu \nu }=\gamma \mathcal{C}^{\ast \nu \alpha }=\gamma
C^{\ast \nu }=0,  \label{tv53a}
\end{equation}%
\begin{equation}
\delta t_{\lambda \mu \nu \vert \alpha }=\delta A_{\mu }=\delta \eta
_{\lambda \mu \nu }=\delta \mathcal{G}_{\mu \nu \vert \alpha }=\delta \eta
=0,  \label{tv54}
\end{equation}%
\begin{equation}
\delta C_{\mu \nu }=\delta \mathcal{C}_{\nu \alpha }=\delta C_{\nu }=0,
\label{tv54a}
\end{equation}%
\begin{equation}
\delta t^{\ast \lambda \mu \nu \vert \alpha }=T^{\lambda \mu \nu \vert
\alpha },\quad \delta A^{\ast \mu }=-\partial _{\nu }F^{\nu \mu },\quad
\delta \eta ^{\ast \lambda \mu \nu }=-4\partial _{\alpha }t^{\ast \lambda
\mu \nu \vert \alpha },  \label{tv55}
\end{equation}%
\begin{equation}
\delta \mathcal{G}^{\ast \mu \nu \vert \alpha }=-\partial _{\lambda }\left(
3t^{\ast \lambda \mu \nu \vert \alpha }-t^{\ast \mu \nu \alpha \vert \lambda
}\right) ,\quad \delta \eta ^{\ast }=-\partial _{\mu }A^{\ast \mu },
\label{tv56}
\end{equation}%
\begin{equation}
\delta C^{\ast \mu \nu }=3\partial _{\lambda }\left( \mathcal{G}^{\ast \mu
\nu \vert \lambda }-\frac{1}{2}\eta ^{\ast \lambda \mu \nu }\right) ,\quad
\delta \mathcal{C}^{\ast \nu \alpha }=\partial _{\mu }\mathcal{G}^{\ast \mu
\left( \nu \vert \alpha \right) },  \label{tv57}
\end{equation}%
\begin{equation}
\delta C^{\ast \nu }=6\partial _{\mu }\left( \mathcal{C}^{\ast \mu \nu }-%
\frac{1}{3}C^{\ast \mu \nu }\right) ,  \label{tv58}
\end{equation}%
where $T^{\lambda \mu \nu \vert \alpha }$ is given in (\ref{tv17}). By
convention, we take $\delta $ and $\gamma $ to act like right derivations.
We note that the action of the Koszul-Tate differential on the antifields
with the antighost numbers equal to two and three gains a simpler expression
if we perform the changes of variables
\begin{equation}
\mathcal{G}^{\prime \ast \mu \nu \vert \vert \alpha }=\mathcal{G}^{\ast \mu
\nu \vert \alpha }+\frac{1}{4}\eta ^{\ast \mu \nu \alpha },\quad \mathcal{C}%
^{\prime \ast \nu \alpha }=\mathcal{C}^{\ast \nu \alpha }-\frac{1}{3}C^{\ast
\nu \alpha },  \label{tv58a}
\end{equation}%
where $\mathcal{G}^{\prime \ast \mu \nu \vert \vert \alpha }$ is still
antisymmetric in its first two indices (but no longer fulfils the identity $%
\mathcal{G}^{\prime \ast \left[ \mu \nu \vert \vert \alpha \right] }\equiv 0$%
) and $\mathcal{C}^{\prime \ast \nu \alpha }$ has no definite symmetry or
antisymmetry properties. With the help of (\ref{tv55})--(\ref{tv58}), we
find that $\delta $ acts on the transformed antifields through the relations
\begin{equation}
\delta \mathcal{G}^{\prime \ast \mu \nu \vert \vert \alpha }=-3\partial
_{\lambda }t^{\ast \lambda \mu \nu \vert \alpha },\quad \delta \mathcal{C}%
^{\prime \ast \nu \alpha }=2\partial _{\mu }\mathcal{G}^{\prime \ast \mu \nu
\vert \vert \alpha },\quad \delta C^{\ast \nu }=6\partial _{\mu }\mathcal{C}%
^{\prime \ast \mu \nu }.  \label{tv58b}
\end{equation}%
The same observation is valid with respect to $\gamma $ if we make the
changes
\begin{equation}
\mathcal{G}_{\mu \nu \vert \vert \alpha }^{\prime }=\mathcal{G}_{\mu \nu
\vert \alpha }+4\eta _{\mu \nu \alpha },\;\mathcal{C}_{\nu \alpha }^{\prime
}=\mathcal{C}_{\nu \alpha }-3C_{\nu \alpha },  \label{tv58ba}
\end{equation}%
in terms of which we can write
\begin{equation}
\gamma t_{\lambda \mu \nu \vert \alpha }=-\frac{1}{4}\partial _{\left[
\lambda \right. }\mathcal{G}_{\left. \mu \nu \vert \vert \alpha \right]
}^{\prime }+\partial _{\left[ \lambda \right. }\mathcal{G}_{\left. \mu \nu %
\right] \vert \vert \alpha }^{\prime },\;\gamma \mathcal{G}_{\mu \nu \vert
\vert \alpha }^{\prime }=\partial _{\left[ \mu \right. }\mathcal{C}_{\left.
\nu \right] \alpha }^{\prime },\;\gamma \mathcal{C}_{\nu \alpha }^{\prime
}=-6\partial _{\nu }C_{\alpha }.  \label{tv58bd}
\end{equation}%
Again, $\mathcal{G}_{\mu \nu \vert \vert \alpha }^{\prime }$ is
antisymmetric in its first two indices, but does not satisfy the identity $%
\mathcal{G}_{\left[ \mu \nu \vert \vert \alpha \right] }^{\prime }\equiv 0$,
while $\mathcal{C}_{\nu \alpha }^{\prime }$ has no definite symmetry or
antisymmetry. We have deliberately chosen the same notations for the
transformed variables (\ref{tv58a}) and (\ref{tv58ba}) since they actually
form pairs that are conjugated in the antibracket.

The Lagrangian BRST differential admits a canonical action in a structure
named antibracket and defined by decreeing the fields/ghosts conjugated with
the corresponding antifields, $s\cdot =\left( \cdot ,S\right) $, where $%
\left( ,\right) $ signifies the antibracket and $S$ denotes the canonical
generator of the BRST symmetry. It is a bosonic functional of ghost number
zero (involving both field/ghost and antifield spectra) that obeys the
master equation
\begin{equation}
\left( S,S\right) =0.  \label{tv59}
\end{equation}%
The master equation is equivalent with the second-order nilpotency of $s$,
where its solution $S$ encodes the entire gauge structure of the associated
theory. Taking into account the formulas (\ref{tv49})--(\ref{tv58}) as well
as the standard actions of $\delta $ and $\gamma $ in canonical form we find
that the complete solution to the master equation for the free model under
study is given by
\begin{eqnarray}
S &=&S_{0}\left[ t_{\lambda \mu \nu \vert \alpha },A_{\mu }\right] +\int
d^{D}x\left[ t^{\ast \lambda \mu \nu \vert \alpha }\left( 3\partial _{\alpha
}\eta _{\lambda \mu \nu }+\partial _{\left[ \lambda \right. }\eta _{\left.
\mu \nu \right] \alpha }+\partial _{\left[ \lambda \right. }\mathcal{G}%
_{\left. \mu \nu \right] \vert \alpha }\right) \right.  \notag \\
&&-\frac{1}{2}\eta ^{\ast \lambda \mu \nu }\partial _{\left[ \lambda \right.
}C_{\left. \mu \nu \right] }+\mathcal{G}^{\ast \mu \nu \vert \alpha }\left(
2\partial _{\alpha }C_{\mu \nu }-\partial _{\left[ \mu \right. }C_{\left.
\nu \right] \alpha }+\partial _{\left[ \mu \right. }\mathcal{C}_{\left. \nu %
\right] \alpha }\right)  \notag \\
&&\left. +C^{\ast \mu \nu }\partial _{\left[ \mu \right. }C_{\left. \nu %
\right] }-3\mathcal{C}^{\ast \nu \alpha }\partial _{\left( \nu \right.
}C_{\left. \alpha \right) }+A^{\ast \mu }\partial _{\mu }\eta \right] ,
\label{tv60}
\end{eqnarray}%
such that it contains pieces with the antighost number ranging from zero to
three.

\section{Brief review of the deformation procedure}

There are three main types of consistent interactions that can be added to a
given gauge theory: The first type deforms only the Lagrangian action, but
not its gauge transformations. The second kind modifies both the action and
its transformations, but not the gauge algebra. The third, and certainly
most interesting category, changes everything, namely, the action, its gauge
symmetries, and the accompanying algebra.

The reformulation of the problem of consistent deformations of a given
action and of its gauge symmetries in the antifield-BRST setting is based on
the observation that if a deformation of the classical theory can be
consistently constructed, then the solution to the master equation for the
initial theory can be deformed into the solution of the master equation for
the interacting theory
\begin{equation}
\bar{S}=S+gS_{1}+g^{2}S_{2}+O\left( g^{3}\right) ,\quad \varepsilon \left(
\bar{S}\right) =0,\quad \mathrm{gh}\left( \bar{S}\right) =0,  \label{tv61}
\end{equation}%
such that
\begin{equation}
\left( \bar{S},\bar{S}\right) =0.  \label{tv62}
\end{equation}%
Here and in the sequel $\varepsilon \left( F\right) $ denotes the Grassmann
parity of $F$. The projection of (\ref{tv61}) on the various powers of the
coupling constant induces the following tower of equations:
\begin{eqnarray}
g^{0} &:&\left( S,S\right) =0,  \label{tv63} \\
g^{1} &:&\left( S_{1},S\right) =0,  \label{tv64} \\
g^{2} &:&\frac{1}{2}\left( S_{1},S_{1}\right) +\left( S_{2},S\right) =0,
\label{tv65} \\
g^{3} &:&\left( S_{1},S_{2}\right) +\left( S_{3},S\right) =0,  \label{tv66}
\\
g^{4} &:&\frac{1}{2}\left( S_{2},S_{2}\right) +\left( S_{1},S_{3}\right)
+\left( S_{4},S\right) =0,  \label{tv66a} \\
&&\vdots  \notag
\end{eqnarray}%
The first equation is satisfied by hypothesis. The second equation
governs the first-order deformation of the solution to the master
equation ($S_{1}$)
and it shows that $S_{1}$ is a BRST co-cycle, $sS_{1}=0$. This means that $%
S_{1}$ pertains to the ghost number zero cohomological space of $s$, $%
H^{0}\left( s\right) $, which is generically non-empty because it is
isomorphic to the space of physical observables of the free theory.
The remaining equations are responsible for the higher-order
deformations of the solution to the master equation. No obstructions
arise in finding solutions to them as long as no further
restrictions, such as spatiotemporal locality, are imposed.
Obviously, only nontrivial first-order deformations should be
considered, since trivial ones ($S_{1}=sB$) lead to trivial
deformations of the initial theory and can be eliminated by
convenient redefinitions of the fields. Ignoring the trivial
deformations, it follows that $S_{1}$ is a nontrivial
BRST-observable, $S_{1}\in H^{0}\left( s\right) $. Once that the
deformation equations (\ref{tv64})--(\ref{tv66a}), etc., have been
solved by means of specific cohomological techniques, from the
consistent nontrivial deformed solution to the master equation one
can extract all the information on the gauge structure of the
resulting interacting theory.

\section{Main results\label{mainres}}

The aim of this paper is to investigate the consistent interactions
that can be added to the action (\ref{tv1}) without modifying either
the field/ghost/antifield spectrum or the number of independent
gauge symmetries. This matter is addressed in the context of the
antifield-BRST deformation procedure described in the above and
relies on computing the solutions to the Eqs.
(\ref{tv64})--(\ref{tv66a}), etc., from the cohomology of the BRST
differential. For obvious reasons, we consider only smooth, local,
and manifestly covariant deformations and, meanwhile, restrict to
Poincar\'{e}-invariant quantities, i.e. we do not allow explicit
dependence on the spatiotemporal coordinates. The smoothness of
deformations refers to the fact that the deformed solution to the
master equation, (\ref{tv61}), is
smooth in the coupling constant $g$ and reduces to the original solution (%
\ref{tv60}) in the free limit ($g=0$). Moreover, we ask that the
deformed gauge theory preserves the Cauchy order of the uncoupled
model, which enforces the requirement that the interacting
Lagrangian is of maximum order equal to two in the spatiotemporal
derivatives of the fields at each order in the coupling constant.
Here, we present the main result without insisting on the cohomology
tools required by the technique of consistent deformations. All
cohomological proofs that lead to the main result are assembled in
two appendices. The first one deals with the construction of the
general form of the first-order deformation of the solution to the
classical master equation and the second investigates the
higher-order deformations. As it is shown in the end of Appendix
\ref{b}, there appear two distinct solutions to (\ref{tv62}) that
exclude each other.

The first type of deformed solution to the master equation
(\ref{tv62}) that is consistent to all orders in the coupling
constant stops at order one
in the coupling constant and reads as%
\begin{equation}
\bar{S}=S+\frac{g}{3\cdot 4!}\int d^{5}x\,\varepsilon ^{\lambda \mu \nu \rho
\kappa }F_{\lambda \mu }F_{\nu \rho }A_{\kappa },  \label{tv117}
\end{equation}%
where $S$ is given in (\ref{tv60}) in $D=5$. This case is not interesting
since it provides no cross-couplings between the vector field and the tensor
field with the mixed-symmetry $\left( 3,1\right) $. It simply restricts the
free Lagrangian action (\ref{tv1}) to evolve on a five-dimensional
space-time and adds to it the second term on the right-hand side of (\ref%
{tv117}), without changing the original gauge transformations (\ref{tv7a})--(%
\ref{tv7b}) and, in consequence, neither the original Abelian gauge
algebra nor the reducibility structure.

The second type of full deformed solution to the master equation (\ref%
{tv62}) ends at order two in the coupling constant and is given by%
\begin{eqnarray}
\bar{S} &=&S+g\int d^{5}x\,\varepsilon ^{\lambda \mu \nu \rho \kappa }\left(
A_{\lambda }^{\ast }\mathcal{F}_{\mu \nu \rho \kappa }-\frac{2}{3}F_{\lambda
\mu }\partial _{\left[ \xi \right. }t_{\left. \nu \rho \kappa \right]
\vert \theta }\sigma ^{\theta \xi }\right)  \notag \\
&&+\frac{16g^{2}}{3}\int d^{5}x\left( \partial _{\left[ \xi \right.
}t_{\left. \nu \rho \kappa \right] \vert \theta }\sigma ^{\theta \xi
}\right)
\partial ^{\left[ \xi ^{\prime }\right. }t^{\left. \nu \rho \kappa \right]
\vert \theta ^{\prime }}\sigma _{\theta ^{\prime }\xi ^{\prime }}.
\label{tv119}
\end{eqnarray}%
We observe that this solution `lives' also in a five-dimensional space-time,
just like the previous one. From (\ref{tv119}) we read all the information
on the gauge structure of the coupled theory. The terms of antighost number
zero in (\ref{tv119}) provide the Lagrangian action. They can be
equivalently organized as%
\begin{equation}
\bar{S}_{0}\left[ t_{\lambda \mu \nu \vert \alpha },A_{\mu }\right] =S_{0}^{%
\mathrm{t}}\left[ t_{\lambda \mu \nu \vert \alpha }\right]
-\frac{1}{4}\int d^{5}x\,\bar{F}_{\mu \nu }\bar{F}^{\mu \nu },
\label{equivdeflag}
\end{equation}%
in terms of the deformed field strength
\begin{equation}
\bar{F}^{\mu \nu }=F^{\mu \nu }+\frac{4g}{3}\varepsilon ^{\mu \nu \alpha
\beta \gamma }\partial _{\left[ \rho \right. }t_{\left. \alpha \beta \gamma %
\right] \vert }^{\;\;\;\;\;\;\;\;\;\rho },  \label{deffstr}
\end{equation}%
where $S_{0}^{\mathrm{t}}\left[ t_{\lambda \mu \nu \vert \alpha
}\right] $ is the Lagrangian action of the massless tensor field
$t_{\lambda \mu \nu \vert \alpha }$
appearing in (\ref{tv1}) in $D=5$. We observe that the action (\ref%
{equivdeflag}) contains only mixing-component terms of order one and two in
the coupling constant. The piece of antighost number one appearing in (\ref%
{tv119}) gives the deformed gauge transformations in the form%
\begin{eqnarray}
\bar{\delta}_{\epsilon ,\chi }t_{\lambda \mu \nu \vert \alpha } &=&-3\partial _{%
\left[ \lambda \right. }\epsilon _{\left. \mu \nu \alpha \right]
}+4\partial _{\left[ \lambda \right. }\epsilon _{\left. \mu \nu
\right] \alpha }+\partial _{\left[ \lambda \right. }\chi _{\left.
\mu \nu \right] \vert \alpha },
\label{defgauge1} \\
\bar{\delta}_{\epsilon ,\chi }A_{\mu } &=&\partial _{\mu }\epsilon
+4g\varepsilon _{\mu \alpha \beta \gamma \delta }\partial ^{\alpha }\epsilon
^{\beta \gamma \delta }.  \label{defgauge2}
\end{eqnarray}%
It is interesting to note that only the gauge transformations of the
vector field are modified during the deformation process. This is
enforced at order one in the coupling constant by a term linear in
the antisymmetrized first-order derivatives of some gauge parameters
from the $(3,1)$ sector. At antighost numbers strictly greater than
one (\ref{tv119}) coincides with the solution (\ref{tv60})
corresponding to the free theory. Consequently, the gauge algebra
and the reducibility structure of the coupled model are not modified
during the deformation procedure, being the same like in the case
of the starting free action (\ref{tv1}) with the gauge transformations (\ref%
{tv7a})--(\ref{tv7b}). It is easy to see from (\ref{equivdeflag}) and (\ref%
{defgauge1})--(\ref{defgauge2}) that if we impose the PT-invariance at the
level of the coupled model, then we obtain no interactions (we must set $g=0$%
\ in these formulas).

It is important to stress that the problem of obtaining consistent
interactions strongly depends on the spatiotemporal dimension. For
instance, if one starts with action (\ref{tv1}) in $D>5$, then one
inexorably gets $\bar{S}=S$, so \emph{no} term can be added to
either the original Lagrangian or its gauge transformations.

\section{Conclusion}

In this paper we have discussed a cohomological approach to the problem of
constructing consistent interactions between a single massless tensor field $%
t_{\lambda \mu \nu \vert \alpha }$ with the mixed symmetry $\left(
3,1\right) $ and one vector field. Under the general assumptions of
smoothness of the deformations in the coupling constant, locality,
(background) Lorentz invariance, Poincar\'{e} invariance, and preservation
of the number of derivatives on each field, we have exhausted all the
consistent, nontrivial couplings. Our final result is rather surprising
since it enables nontrivial cross-couplings between these types of fields in
five dimensions and also allows the deformation of the gauge transformations
of the vector field. Although the cross-couplings break the PT invariance
and are merely mixing-component terms, still this is the first situation
encountered so far where there exist nontrivial deformations involving
mixed-symmetry tensor fields complying with the hypothesis on the
preservation of the number of derivatives on each field. If we relax this
condition, it is possible that other interactions become consistent as well.

\section*{Acknowledgment}

The authors are partially supported by the European Commission FP6 program
MRTN-CT-2004-005104 and by the type A grant 304/2004 with the Romanian
National Council for Academic Scientific Research and the Romanian Ministry
of Education and Research.

\appendix%

\section{First-order deformation}

Here, we determine the most general form of the first-order
deformation of the solution to the master equation that complies
with all the hypotheses exposed at the beginning of Sec.
\ref{mainres} (smoothness in the coupling constant, locality,
Lorentz-covariance, Poincar\'{e} invariance, and preservation of the
Cauchy order of the uncoupled model). In view of this, we initially
compute the main cohomological ingredients necessary at the
construction of the local cohomology of the BRST differential at
ghost number zero. If we make the notation $S_{1}=\int d^{D}x\,a$,
with $a$ a local function, then the local form of the Eq.
(\ref{tv64}), which we have seen that controls the first-order
deformation of the solution to the master equation, becomes
\begin{equation}
sa=\partial _{\mu }m^{\mu },\quad \mathrm{gh}\left( a\right) =0,\quad
\varepsilon \left( a\right) =0,  \label{tv65a}
\end{equation}%
for some local $m^{\mu }$, and it shows that the non-integrated
density of the first-order deformation pertains to the local
cohomology of $s$ at ghost number zero, $a\in H^{0}\left( s\vert
d\right) $, where $d$ denotes the exterior differential in
space-time. In order to analyze the above equation, we develop $a$
according to the antighost number
\begin{equation}
a=\sum\limits_{k=0}^{I}a_{k},\quad \mathrm{agh}\left( a_{k}\right) =k,\quad
\mathrm{gh}\left( a_{k}\right) =0,\quad \varepsilon \left( a_{k}\right) =0,
\label{tv65b}
\end{equation}%
and assume, without loss of generality, that the above decomposition stops
at some finite value of the antighost number, $I$. By taking into account
the splitting (\ref{tv39}) of the BRST differential, the Eq. (\ref{tv65a})
is equivalent to a tower of local equations, corresponding to the different
decreasing values of the antighost number
\begin{eqnarray}
\gamma a_{I} &=&\partial _{\mu }\overset{(I)}{m}^{\mu },  \label{tv65c} \\
\delta a_{I}+\gamma a_{I-1} &=&\partial _{\mu }\overset{(I-1)}{m}^{\mu },
\label{tv65d} \\
\delta a_{k}+\gamma a_{k-1} &=&\partial _{\mu }\overset{(k-1)}{m}^{\mu
},\quad I-1\geq k\geq 1,  \label{tv65e}
\end{eqnarray}%
where $\left( \overset{(k)}{m}^{\mu }\right) _{k=\overline{0,I}}$ are some
local currents with $\mathrm{agh}\left( \overset{(k)}{m}^{\mu }\right) =k$.
It can be proved that we can replace the Eq. (\ref{tv65c}) at strictly
positive antighost numbers with
\begin{equation}
\gamma a_{I}=0,\quad \mathrm{agh}\left( a_{I}\right) =I>0.  \label{tv65f}
\end{equation}%
The proof can be done like in the Appendix A, Corollary 1 from \cite%
{noijhep31}, with the precaution to include in an appropriate manner the
vector field sector. In conclusion, under the assumption that $I>0$, the
representative of highest antighost number from the non-integrated density
of the first-order deformation can always be taken to be $\gamma $-closed,
such that the Eq. (\ref{tv65a}) associated with the local form of the
first-order deformation is completely equivalent to the tower of equations (%
\ref{tv65d})--(\ref{tv65e}) and (\ref{tv65f}).

Before proceeding to the analysis of the solutions to the first-order
deformation equation, we briefly comment on the uniqueness and triviality of
such solutions. Due to the second-order nilpotency of $\gamma $ ($\gamma
^{2}=0$), the solution to the top equation, (\ref{tv65f}), is clearly unique
up to $\gamma $-exact contributions, $a_{I}\rightarrow a_{I}+\gamma b_{I}$.
Meanwhile, if it turns out that $a_{I}$ reduces to $\gamma $-exact terms
only, $a_{I}=\gamma b_{I}$, then it can be made to vanish, $a_{I}=0$. In
other words, the nontriviality of the first-order deformation $a$ is
translated at its highest antighost number component into the requirement
that $a_{I}\in H^{I}\left( \gamma \right) $, where $H^{I}\left( \gamma
\right) $ denotes the cohomology of the exterior longitudinal derivative $%
\gamma $ at pure ghost number equal to $I$. At the same time, the general
condition on the non-integrated density of the first-order deformation to be
in a nontrivial cohomological class of $H^{0}\left( s\vert d\right) $ shows
on the one hand that the solution to (\ref{tv65a}) is unique up to $s$-exact
pieces plus divergences and on the other hand that if the general solution
to (\ref{tv65a}) is found to be completely trivial, $a=sb+\partial _{\mu
}n^{\mu }$, then it can be made to vanish, $a=0$.

\subsection{Basic cohomologies}

In the light of the above discussion, we pass now to the investigation of
the solutions to the Eqs. (\ref{tv65f}) and (\ref{tv65d})--(\ref{tv65e}). We
have seen that the solution to (\ref{tv65f}) belongs to the cohomology of
the exterior longitudinal derivative, such that we need to compute $H\left(
\gamma \right) $ in order to construct the component of highest antighost
number from the first-order deformation. This matter is solved with the help
of the definitions (\ref{tv49})--(\ref{tv53a}). As it has been shown in \cite%
{noijhep31}, the most general, nontrivial representative from $H\left(
\gamma \right) $ in the $t$-sector is written like%
\begin{equation}
a_{I}^{\mathrm{t}}= \alpha _{I}\left( \left[ \Pi ^{\ast (\mathrm{t})\Delta }%
\right] ,\left[ K_{\lambda \mu \nu \xi \vert \alpha \beta }\right] \right)
\omega ^{(\mathrm{t})I}\left( \mathcal{F}_{\lambda \mu \nu \alpha },C_{\nu
}\right) ,\quad I>0,  \label{hgammat}
\end{equation}%
where we employed the notation
\begin{equation}
\mathcal{F}_{\lambda \mu \nu \alpha }\equiv \partial _{\left[ \lambda
\right. }\eta _{\left. \mu \nu \alpha \right] }.  \label{tv67}
\end{equation}%
In the above $\Pi ^{\ast (\mathrm{t})\Delta }$ denote the antifields from
the $t$-sector, the notation $f\left( \left[ q\right] \right) $ means that
the function $f$ depends on the variable $q$ and its subsequent derivatives
up to a finite number, and $\omega ^{(\mathrm{t})I}$ are the elements of
pure ghost number $I$ (and obviously of antighost number zero) of a basis in
the space of polynomials in $\mathcal{F}_{\lambda \mu \nu \alpha }$ and $%
C_{\nu }$, which is finite-dimensional since all these variables
anticommute. (The spatiotemporal derivatives of $\mathcal{F}_{\lambda \mu
\nu \alpha }$ and $C_{\nu }$ have been shown in \cite{noijhep31} to be $%
\gamma $-exact and the same is valid with respect to the derivatives of the
ghosts for ghosts for ghosts $C_{\nu }$. Regarding the ghosts for ghosts $%
C_{\mu \nu }$ and $\mathcal{C}_{\nu \alpha }$, in the same paper we have
proved that there is no linear combination of these undifferentiated ghosts
in $H\left( \gamma \right) $ and all the elements from $H\left( \gamma
\right) $ involving their derivatives are $\gamma $-exact.) At the level of
the vector theory, all the corresponding antifields, the Abelian field
strength (\ref{abfstr}), and their spatiotemporal derivatives are nontrivial
objects from $H^{0}\left( \gamma \right) $, while the non-differentiated
ghost $\eta $ is the sole nontrivial element from $H\left( \gamma \right) $
at strictly positive values of the pure ghost number. (Its spatiotemporal
derivatives of any order are $\gamma $-exact, according to the second
definition in (\ref{tv49}).) Combining (\ref{hgammat}) with the above
argument, we can state that the most general, nontrivial representative from
$H\left( \gamma \right) $ for the overall theory (\ref{tv1}) reads as
\begin{equation}
a_{I}=\alpha _{I}\left( \left[ \Pi ^{\ast \Delta }\right] ,\left[ K_{\lambda
\mu \nu \xi \vert \alpha \beta }\right] ,\left[ F_{\mu \nu }\right] \right)
\omega ^{I}\left( \mathcal{F}_{\lambda \mu \nu \alpha },\eta ,C_{\nu
}\right) ,\quad I>0,  \label{tv81}
\end{equation}%
where here $\Pi ^{\ast \Delta }$ denote all the antifields and $\omega ^{I}$
are now the elements of pure ghost number $I$ (and obviously of antighost
number zero) of a basis in the space of polynomials in $\mathcal{F}_{\lambda
\mu \nu \alpha }$, $\eta $, and $C_{\nu }$ (which is again
finite-dimensional). The objects $\alpha _{I}$ (obviously nontrivial in $%
H^{0}\left( \gamma \right) $) were taken to have a bounded number of
derivatives, and therefore they are polynomials in the antifields
$\Pi ^{\ast \Delta }$, in the curvature tensor $K_{\lambda \mu \nu
\xi \vert \alpha \beta }$, in the field strength $F_{\mu \nu }$, and
in their derivatives. They are required to fulfill the property
$\mathrm{agh}\left( \alpha _{I}\right) =I$ in order to ensure that
the ghost number of $a_{I}$ is equal to zero. Due to their $\gamma
$-closeness, $\alpha _{I}$ will be called ``invariant polynomials".
At zero antighost number, the invariant polynomials are polynomials
in the curvature $K_{\lambda \mu \nu \xi \vert \alpha \beta }$, in
the field strength $F_{\mu \nu } $, and in their derivatives.

Replacing the solution (\ref{tv81}) into the Eq. (\ref{tv65d}) and taking
into account the definitions (\ref{tv54})--(\ref{tv58}), we remark that a
necessary (but not sufficient) condition for the existence of (nontrivial)
solutions $a_{I-1}$ is that the invariant polynomials $\alpha _{I}$ are
(nontrivial) objects from the local cohomology of the Koszul-Tate
differential $H\left( \delta \vert d\right) $ at antighost number $I>0$ and
pure ghost number equal to zero\footnote{\label{local}We recall that the
local cohomology $H\left( \delta \vert d\right) $ is completely trivial at
both strictly positive antighost \textit{and} pure ghost numbers (for
instance, see~\cite{gen1a}, Theorem 5.4, \cite{gen1b}, and~\cite{commun1}).}%
, i.e.,
\begin{equation}
\delta \alpha _{I}=\partial _{\mu }\overset{(I-1)}{j}^{\mu },\quad \mathrm{%
agh}\left( \overset{(I-1)}{j}^{\mu }\right) =I-1\geq 0,\quad \mathrm{pgh}%
\left( \overset{(I-1)}{j}^{\mu }\right) =0.  \label{tv82}
\end{equation}%
The above notation is generic, in the sense that $\alpha _{I}$ and $\overset{%
(I-1)}{j}^{\mu }$ may actually carry supplementary Lorentz indices.
Consequently, we need to investigate some of the main properties of the
local cohomology of the Koszul-Tate differential $H\left( \delta \vert
d\right) $ at pure ghost number zero and strictly positive antighost numbers
in order to fully determine the component $a_{I}$ of highest antighost
number from the first-order deformation. As the free model under study is a
linear gauge theory of Cauchy order equal to four, the general results from~%
\cite{gen1a,gen1b} (also see~\cite{lingr,gen2,multi}) ensure that $H\left(
\delta \vert d\right) $ (at pure ghost number zero) is trivial at antighost
numbers strictly greater than its Cauchy order
\begin{equation}
H_{I}\left( \delta \vert d\right) =0,\;I>4.  \label{tv83}
\end{equation}%
Moreover, if the invariant polynomial $\alpha _{I}$, with $\mathrm{agh}%
\left( \alpha _{I}\right) =I\geq 4$, is trivial in $H_{I}\left( \delta \vert
d\right) $, then it can be taken to be trivial also in $H_{I}^{\mathrm{inv}%
}\left( \delta \vert d\right) $%
\begin{equation}
\left( \alpha _{I}=\delta b_{I+1}+\partial _{\mu }\overset{(I)}{c}^{\mu },\;%
\mathrm{agh}\left( \alpha _{I}\right) =I\geq 4\right) \Rightarrow \alpha
_{I}=\delta \beta _{I+1}+\partial _{\mu }\overset{(I)}{\gamma }^{\mu },
\label{tv83a}
\end{equation}%
with $\beta _{I+1}$ and $\overset{(I)}{\gamma }^{\mu }$ invariant
polynomials. [An element of $H_{I}^{\mathrm{inv}}\left( \delta \vert
d\right) $ is defined via an equation similar to (\ref{tv82}), but with the
corresponding current also an invariant polynomial.] The result (\ref{tv83a}%
) can be proved like in the Appendix B, Theorem 3, from \cite{noijhep31}, up
to the observation that the vector field sector must also be considered.
This is important since together with (\ref{tv83}) ensures that the entire
local cohomology of the Koszul-Tate differential in the space of invariant
polynomials (characteristic cohomology) is trivial in antighost number
strictly greater than four
\begin{equation}
H_{I}^{\mathrm{inv}}\left( \delta \vert d\right) =0,\quad I>4.  \label{tv83b}
\end{equation}%
Looking at the definitions (\ref{tv58b}) involving the transformed
antifields (\ref{tv58a}) and taking into account the formulas (\ref{tv54})--(%
\ref{tv56}) with respect to the vector field theory, we can organize the
nontrivial representatives of $H_{I}\left( \delta \vert d\right) $ (at pure
ghost number equal to zero) and of $\left( H_{I}^{\mathrm{inv}}\left( \delta
\vert d\right) \right) _{I\geq 2}$ in the following array
\begin{equation}
\begin{array}{cc}
\mathrm{agh} &
\begin{array}{l}
\mathrm{nontrivial\;representatives} \\
\mathrm{spanning\;}H_{I}\left( \delta \vert d\right) \;\mathrm{and}\;H_{I}^{%
\mathrm{inv}}\left( \delta \vert d\right)%
\end{array}
\\
I>4 & \mathrm{none} \\
I=4 & C^{\ast \nu } \\
I=3 & \mathcal{C}^{\prime \ast \nu \alpha } \\
I=2 & \mathcal{G}^{\prime \ast \mu \nu \vert \alpha },\eta ^{\ast }%
\end{array}%
.  \label{tvabledelta}
\end{equation}%
We remark that in $\left( H_{I}\left( \delta \vert d\right) \right) _{I\geq
2}$ or $\left( H_{I}^{\mathrm{inv}}\left( \delta \vert d\right) \right)
_{I\geq 2}$ there is no nontrivial element that effectively involves the
curvature tensor $K_{\lambda \mu \nu \xi \vert \alpha \beta }$, the field
strength $F_{\mu \nu }$, and/or their derivatives, and the same stands for
the quantities that are more than linear in the antifields and/or depend on
their derivatives. It is also important to note that the vector field sector
brings no contribution to $\left( H_{I}\left( \delta \vert d\right) \right)
_{I>2} $. In contrast to the groups $\left( H_{I}\left( \delta \vert
d\right) \right) _{I\geq 2}$ and $\left( H_{I}^{\mathrm{inv}}\left( \delta
\vert d\right) \right) _{I\geq 2}$, which are finite-dimensional, the
cohomology $H_{1}\left( \delta \vert d\right) $ at pure ghost number zero,
that is related to global symmetries and ordinary conservation laws, is
infinite-dimensional since the theory is free.

The previous results on $H\left( \delta \vert d\right) $ and $H^{\mathrm{inv}%
}\left( \delta \vert d\right) $ at strictly positive antighost numbers are
important because they control the obstructions to removing the antifields
from the first-order deformation. Indeed, due to (\ref{tv83b}), it follows
that we can successively eliminate all the pieces with $I>4$ from the
non-integrated density of the first-order deformation by adding only trivial
terms (the proof is similar to that from the Appendix C in \cite{noijhep31}
modulo the inclusion of the vector field sector), so we can take, without
loss of nontrivial objects, the condition $I\leq 4$ in the decomposition (%
\ref{tv65b}). The last representative is of the form (\ref{tv81}), where the
invariant polynomial is necessarily a nontrivial object from $H_{I}^{\mathrm{%
inv}}\left( \delta \vert d\right) $ for $I=2,3,4$ and respectively from $%
H_{1}\left( \delta \vert d\right) $ for $I=1$.

\subsection{Computation of first-order deformations}

Now, we have at hand all the necessary ingredients for computing the general
form of the first-order deformation of the solution to the master equation
as solution to the local equation (\ref{tv65a}). In view of this, we
decompose the first-order deformation like%
\begin{equation}
a=a^{\mathrm{t}}+a^{\mathrm{A}}+a^{\mathrm{t-A}},  \label{tv84}
\end{equation}%
where $a^{\mathrm{t}}$ denotes the part responsible for the
self-interactions of the field $t_{\lambda \mu \nu \vert \alpha }$, $a^{%
\mathrm{A}}$ is related to the deformations of the vector field, and $a^{%
\mathrm{t-A}}$ signifies the component that describes only the
cross-couplings between $t_{\lambda \mu \nu \vert \alpha }$ and the vector
field. Obviously, the Eq. (\ref{tv65a}) becomes equivalent with three
equations, one for each component%
\begin{equation}
sa^{\mathrm{t}}=\partial _{\mu }m_{\mathrm{t}}^{\mu },\quad sa^{\mathrm{A}%
}=\partial _{\mu }m_{\mathrm{A}}^{\mu },\quad sa^{\mathrm{t-A}}=\partial
_{\mu }m_{\mathrm{t-A}}^{\mu }.  \label{tv85c}
\end{equation}%
The solutions to the first equation have been studied in \cite{noijhep31}
and were proved to be trivial%
\begin{equation}
a^{\mathrm{t}}=0.  \label{tv86a}
\end{equation}%
The solutions to the second equation have been investigated in \cite%
{citatiicupforme1,citatiicupforme2}. If we select among them only
the nontrivial solutions in $D\geq 5$ space-time dimensions
containing at most two derivatives of the vector field, then we are
left with one candidate living
precisely in $D=5$%
\begin{equation}
a^{\mathrm{A}}=\frac{c}{3\cdot 4!}\delta _{D5}\varepsilon ^{\lambda \mu \nu
\rho \kappa }F_{\lambda \mu }F_{\nu \rho }A_{\kappa },  \label{tv86b}
\end{equation}%
with $c$ an arbitrary, real constant.

In order to solve the third equation in (\ref{tv85c}) we decompose $a^{%
\mathrm{t-A}}$ along the antighost number like in (\ref{tv65b}) and stop at $%
I=4$%
\begin{equation}
a^{\mathrm{t-A}}=a_{0}^{\mathrm{t-A}}+a_{1}^{\mathrm{t-A}}+a_{2}^{\mathrm{t-A%
}}+a_{3}^{\mathrm{t-A}}+a_{4}^{\mathrm{t-A}},  \label{tv87}
\end{equation}%
where $a_{4}^{\mathrm{t-A}}$ can be taken as solution to the equation $%
\gamma a_{4}^{\mathrm{t-A}}=0$, and therefore it is of the form (\ref{tv81})
for $I=4$, with $\alpha _{4}$ an invariant polynomial from $H_{4}^{\mathrm{%
inv}}\left( \delta \vert d\right) $. Because $H_{4}^{\mathrm{inv}}\left(
\delta \vert d\right) $ is spanned by $C^{\ast \nu }$ (see (\ref{tvabledelta}%
)) and $a_{4}^{\mathrm{t-A}}$ must yield cross-couplings between $t_{\lambda
\mu \nu \vert \alpha }$ and $A_{\mu }$ with maximum two spatiotemporal
derivatives, it follows that the eligible basis elements at pure ghost
number equal to four remain $\omega _{\mu }^{4}\left( \mathcal{F}_{\lambda
\mu \nu \alpha },\eta ,C_{\nu }\right) =C_{\mu }\eta $, so we have (up to
trivial, $\gamma $-exact contributions) that%
\begin{equation}
a_{4}^{\mathrm{t-A}}=c_{1}C^{\ast \mu }C_{\mu }\eta .  \label{tv88}
\end{equation}%
Replacing $a_{4}^{\mathrm{t-A}}$ into an equation similar to (\ref{tv65d})
for $I=4$ and computing $\delta a_{4}^{\mathrm{t-A}}$, it follows that%
\begin{equation}
a_{3}^{\mathrm{t-A}}=c_{1}\mathcal{C}^{\prime \ast \mu \nu }\left( \mathcal{C%
}_{\mu \nu }^{\prime }\eta +6A_{\mu }C_{\nu }\right) .  \label{tv89}
\end{equation}%
In the right-hand side of (\ref{tv89}) one should add $\bar{a}_{3}^{\mathrm{%
t-A}}$ as solution to the `homogeneous' equation $\gamma \bar{a}_{3}^{%
\mathrm{t-A}}=0$. In this particular case one has $\bar{a}_{3}^{\mathrm{t-A}%
}=0$ because there are no basis elements $\omega ^{3}\left( \mathcal{F}%
_{\lambda \mu \nu \alpha },\eta ,C_{\nu }\right) $ enabling cross-couplings
with at most two derivatives in the corresponding $a_{0}^{\mathrm{t-A}}$.
Indeed, such elements must include precisely one fermionic ghost $\eta $
since $H_{3}^{\mathrm{inv}}\left( \delta \vert d\right) $ is spanned only by
antifields from the $t$-sector, and therefore they are forced to be
quadratic in the combination $\mathcal{F}_{\lambda \mu \nu \alpha }$ defined
in (\ref{tv67}). A simple estimation shows that even if consistent, the
associated $a_{0}^{\mathrm{t-A}}$ would contain three derivatives of the
fields, which is unacceptable. Acting with $\delta $ on (\ref{tv89}) we
arrive at%
\begin{equation}
\delta a_{3}^{\mathrm{t-A}}=\gamma \left[ -c_{1}\mathcal{G}^{\prime \ast \mu
\nu \vert \vert \alpha }\left( \mathcal{G}_{\mu \nu \vert \vert \alpha
}^{\prime }\eta -A_{\left[ \mu \right. }\mathcal{C}_{\left. \nu \right]
\alpha }^{\prime }\right) \right] +6c_{1}\mathcal{G}^{\prime \ast \mu \nu
\vert \vert \alpha }F_{\mu \nu }C_{\alpha }+\partial _{\mu }\overset{(2)}{z}%
_{\mathrm{t-A}}^{\mu }.  \label{tv90}
\end{equation}%
Comparing (\ref{tv90}) with an equation of the form (\ref{tv65e})
for $k=3$, we infer that $a_{2}^{\mathrm{t-A}}$ exists if and only
if the second term in the right-hand side of (\ref{tv90}) can be
written in a $\gamma $-exact modulo $d$ form. However, this is
impossible since this term is a nontrivial element from $H^{3}\left(
\gamma \right) $ that cannot be represented in a
divergence-like form. Consequently, we must set $c_{1}=0$ in (\ref{tv88})--(%
\ref{tv89}), so $a^{\mathrm{t-A}}$ can stop earliest at antighost number two.

In this case we have that%
\begin{equation}
a^{\mathrm{t-A}}=a_{0}^{\mathrm{t-A}}+a_{1}^{\mathrm{t-A}}+a_{2}^{\mathrm{t-A%
}}.  \label{tv91}
\end{equation}%
Here, $a_{2}^{\mathrm{t-A}}$ is solution to the equation $\gamma a_{2}^{%
\mathrm{t-A}}=0$, and thus of the type (\ref{tv81}) for $I=2$, with $\alpha
_{2}$ an invariant polynomial from $H_{2}^{\mathrm{inv}}\left( \delta \vert
d\right) $. The basis elements $\omega ^{2}$ at pure ghost number two are in
this case spanned by $\mathcal{F}_{\lambda \mu \nu \alpha }\mathcal{F}%
_{\lambda \mu \nu \alpha }$ and $\mathcal{F}_{\lambda \mu \nu \alpha }\eta $%
. The former is forbidden by the hypothesis on the maximum number of
derivatives in $a_{0}^{\mathrm{t-A}}$ being equal to two (if consistent,
then it would produce an $a_{0}^{\mathrm{t-A}}$ with three derivatives), so
only the latter is allowed. Inspecting next the formula (\ref{tvabledelta})
at $I=2$, we remark that although this is the first place where an antifield
from the vector sector may appear, it is impossible to construct a Lorentz
scalar in $D\geq 5$ dimensions by `gluing' $\eta ^{\ast }$ to $\mathcal{F}%
_{\lambda \mu \nu \alpha }\eta $. In this way we remain with two possible
pieces in $a_{2}^{\mathrm{t-A}}$%
\begin{equation}
a_{2}^{\mathrm{t-A}}=\left( \delta _{D7}c_{2}\varepsilon _{\lambda \mu \nu
\alpha \beta \gamma \delta }\mathcal{G}^{\prime \ast \lambda \mu \vert \vert
\nu }+\delta _{D5}c_{3}\varepsilon _{\lambda \alpha \beta \gamma \delta }%
\mathcal{G}^{\prime \ast \lambda \mu \vert \vert \nu }\sigma _{\mu \nu
}\right) \mathcal{F}^{\alpha \beta \gamma \delta }\eta ,  \label{tv92}
\end{equation}%
where $\delta _{Dn}$ is the Kronecker symbol. Substituting the above $a_{2}^{%
\mathrm{t-A}}$ into an equation similar to (\ref{tv65d}) for $I=2$ and
computing $\delta a_{2}^{\mathrm{t-A}}$, it results that%
\begin{eqnarray}
a_{1}^{\mathrm{t-A}} &=&\left( \delta _{D7}\frac{c_{2}}{3}\varepsilon
_{\lambda \mu \nu \alpha \beta \gamma \delta }t^{\ast \lambda \mu \nu \vert
\theta }+\delta _{D5}c_{3}\varepsilon _{\lambda \alpha \beta \gamma \delta
}t^{\ast \theta \lambda }\right) \left[ \left( \partial ^{\left[ \alpha
\right. }t^{\left. \beta \gamma \delta \right] \vert \rho }\right) \sigma
_{\theta \rho }\eta \right.  \notag \\
&&\left. -3A_{\theta }\mathcal{F}^{\alpha \beta \gamma \delta }\right] +\bar{%
a}_{1}^{\mathrm{t-A}},  \label{tv93}
\end{eqnarray}%
where $\bar{a}_{1}^{\mathrm{t-A}}$ is the general solution to the
`homogeneous' equation $\gamma \bar{a}_{1}^{\mathrm{t-A}}=0$. The solution $%
\bar{a}_{1}^{\mathrm{t-A}}$ requires special attention since although it is
still of the form (\ref{tv81}) with $I=1$, the corresponding invariant
polynomial $\alpha _{1}$ is no longer restricted to belong to $H_{1}^{%
\mathrm{inv}}\left( \delta \vert d\right) $; it pertains to the larger,
infinite-dimensional space $H_{1}\left( \delta \vert d\right) $. However, if
we select from (\ref{tv81}) with $I=1$ the representatives that comply with
all the requirements, like the restrictions on the dimension of the
space-time, on the maximum number of derivatives in the corresponding $%
a_{0}^{\mathrm{t-A}}$, and on the interaction vertices to generate
cross-couplings between the two types of fields (and not self-interactions),
then we obtain just two independent contributions%
\begin{equation}
\bar{a}_{1}^{\mathrm{t-A}}=c_{4}t^{\ast \lambda \mu }F_{\lambda \mu }\eta
+\delta _{D5}c_{5}\varepsilon _{\lambda \alpha \beta \gamma \delta }A^{\ast
\lambda }\mathcal{F}^{\alpha \beta \gamma \delta },  \label{tv94}
\end{equation}%
where $F_{\lambda \mu }$ is the field strength of the vector field, (\ref%
{abfstr}). Applying $\delta $ on (\ref{tv93}) and taking into account (\ref%
{tv94}) we arrive at
\begin{eqnarray}
&&\delta a_{1}^{\mathrm{t-A}}=-\gamma \left( \frac{2c_{5}}{3}\varepsilon
_{\mu \nu \alpha \beta \gamma }F^{\mu \nu }\partial ^{\left[ \rho \right.
}t^{\left. \alpha \beta \gamma \right] \vert \theta }\sigma _{\theta \rho
}+a_{0}^{\mathrm{t-A}}\right) +\partial _{\mu }\overset{(0)}{z}_{\mathrm{t-A}%
}^{\mu }  \notag \\
&&+\left( \delta _{D7}\frac{c_{2}}{3}\varepsilon _{\lambda \mu \nu \alpha
\beta \gamma \delta }T^{\lambda \mu \nu \vert \theta }+\delta
_{D5}c_{3}\varepsilon _{\lambda \alpha \beta \gamma \delta }T^{\theta
\lambda }\right) \left[ \left( \partial ^{\left[ \alpha \right. }t^{\left.
\beta \gamma \delta \right] \vert \rho }\right) \sigma _{\theta \rho }\eta
\right.  \notag \\
&&\left. -3A_{\theta }\mathcal{F}^{\alpha \beta \gamma \delta }\right]
+c_{4}T^{\lambda \mu }F_{\lambda \mu }\eta .  \label{tv95}
\end{eqnarray}%
Comparing (\ref{tv95}) with an equation similar to (\ref{tv65e}) for $k=1$
it is now clear that $a_{0}^{\mathrm{t-A}}$ exists if and only if the last
terms in the right-hand side of (\ref{tv95}) can be written in a $\gamma $%
-exact modulo $d$ form. Moreover, the terms proportional with $c_{2}$, $%
c_{3} $, and respectively $c_{4}$ must individually satisfy a $\gamma $%
-exactness modulo $d$ condition. This is because the first two kinds of
terms involve Levi-Civita symbols in different dimensions, the third type
contains no such symbol, and the definitions (\ref{tv49})--(\ref{tv53a}) of $%
\gamma $ acting on the fields/ghosts are also free of Levi-Civita symbols.
This means that the equations%
\begin{eqnarray}
&&\delta _{D7}\frac{c_{2}}{3}\varepsilon _{\lambda \mu \nu \alpha \beta
\gamma \delta }T^{\lambda \mu \nu \vert \theta }\left[ \left( \partial ^{%
\left[ \alpha \right. }t^{\left. \beta \gamma \delta \right] \vert \rho
}\right) \sigma _{\theta \rho }\eta -3A_{\theta }\mathcal{F}^{\alpha \beta
\gamma \delta }\right] =  \notag \\
&&\gamma a_{0c_{2}}^{\mathrm{t-A}}-\partial _{\mu }\overset{(0)}{m}_{\mathrm{%
t-A}c_{2}}^{\mu },  \label{tv96a}
\end{eqnarray}%
\begin{equation}
\delta _{D5}c_{3}\varepsilon _{\lambda \alpha \beta \gamma \delta }T^{\theta
\lambda }\left[ \left( \partial ^{\left[ \alpha \right. }t^{\left. \beta
\gamma \delta \right] \vert \rho }\right) \sigma _{\theta \rho }\eta
-3A_{\theta }\mathcal{F}^{\alpha \beta \gamma \delta }\right] =\gamma
a_{0c_{3}}^{\mathrm{t-A}}-\partial _{\mu }\overset{(0)}{m}_{\mathrm{t-A}%
c_{3}}^{\mu },  \label{tv96b}
\end{equation}%
\begin{equation}
c_{4}T^{\lambda \mu }F_{\lambda \mu }\eta =\gamma a_{0c_{4}}^{\mathrm{t-A}%
}-\partial _{\mu }\overset{(0)}{m}_{\mathrm{t-A}c_{4}}^{\mu },  \label{tv96c}
\end{equation}%
are necessary and sufficient in order to ensure the existence of $a_{0}^{%
\mathrm{t-A}}$. However, it can be shown that neither of the quantities from
the left-hand sides of (\ref{tv96a})--(\ref{tv96c}) can be set in a $\gamma $%
-exact modulo $d$ form. The proofs are given below.

We assume that $a_{0c_{2}}^{\mathrm{t-A}}$ as solution to the Eq. (\ref%
{tv96a}) exists. Then, from the left-hand of this equation combined with the
formula (\ref{tv49}) and the second relation from (\ref{tv50}) it follows
that it can be represented as a sum of terms, each of them being linear in
the vector field, quadratic in $t_{\lambda \mu \nu \vert \alpha }$, and
containing exactly two spatiotemporal derivatives. Up to irrelevant, total
divergences, we can always move the derivatives such as to act only on the $%
t $-fields%
\begin{equation}
a_{0c_{2}}^{\mathrm{t-A}}=c_{2}\delta _{D7}A_{\mu }f_{\mathrm{lin}}^{\mu
}\left( \partial t\partial t,t\partial \partial t\right) ,  \label{tv96a1}
\end{equation}%
where $f_{\mathrm{lin}}^{\mu }$ is linear in its arguments and contains one
seven-dimensional Levi-Civita symbol. Acting with $\gamma $ on (\ref{tv96a1}%
), we deduce%
\begin{eqnarray}
\gamma a_{0c_{2}}^{\mathrm{t-A}} &=&-c_{2}\delta _{D7}\left[ \partial _{\mu
}f_{\mathrm{lin}}^{\mu }\left( \partial t\partial t,t\partial \partial
t\right) \right] \eta +c_{2}\delta _{D7}A_{\mu }\gamma f_{\mathrm{lin}}^{\mu
}\left( \partial t\partial t,t\partial \partial t\right)  \notag \\
&&+\partial _{\mu }\left[ c_{2}\delta _{D7}f_{\mathrm{lin}}^{\mu }\left(
\partial t\partial t,t\partial \partial t\right) \eta \right] .
\label{tv96a2}
\end{eqnarray}%
Comparing the first term from the right-hand side of (\ref{tv96a2}) with the
first term from the left-hand side of (\ref{tv96a}), we obtain a necessary
condition for the existence of $a_{0c_{2}}^{\mathrm{t-A}}$%
\begin{equation}
\varepsilon _{\lambda \mu \nu \alpha \beta \gamma \delta }T^{\lambda \mu \nu
\vert \theta }\partial ^{\left[ \alpha \right. }t^{\left. \beta \gamma
\delta \right] \vert \rho }\sigma _{\theta \rho }=\partial _{\mu }M^{\mu }.
\label{tv96a3}
\end{equation}%
By direct computation we have that%
\begin{eqnarray}
&&\varepsilon _{\lambda \mu \nu \alpha \beta \gamma \delta }T^{\lambda \mu
\nu \vert \theta }\partial ^{\left[ \alpha \right. }t^{\left. \beta \gamma
\delta \right] \vert \rho }\sigma _{\theta \rho }=-2\varepsilon _{\lambda
\mu \nu \alpha \beta \gamma \delta }t^{\lambda \mu \nu \vert [\theta ,\xi
]}\partial ^{\alpha }t_{\quad \ [\theta ,\xi ]}^{\beta \gamma \delta \vert }
\notag \\
&&+\partial _{\mu }\left( 4\varepsilon _{\lambda \xi \nu \alpha \beta \gamma
\delta }t^{\lambda \xi \nu \vert [\theta ,\mu ]}\partial ^{\left[ \alpha
\right. }t^{\beta \gamma \delta \vert \rho }\sigma _{\theta \rho }\right) ,
\label{tv96a4}
\end{eqnarray}%
where we used the generic notation $k_{,\mu }=\partial k/\partial x^{\mu }$.
Obviously, the condition (\ref{tv96a3}) does not hold since the first term
in the right-hand side of (\ref{tv96a4}) does not reduce to a total
divergence. As a consequence, (\ref{tv96a}) is satisfied only for the
trivial choice $c_{2}=0$. Passing to the next equation, (\ref{tv96b}), an
absolutely similar argument leads to the conclusion that a necessary
condition for the existence of $a_{0c_{3}}^{\mathrm{t-A}}$ is
\begin{equation}
\varepsilon _{\lambda \alpha \beta \gamma \delta }T^{\theta \lambda
}\partial ^{\left[ \alpha \right. }t^{\left. \beta \gamma \delta \right]
\vert \rho }\sigma _{\theta \rho }=\partial _{\mu }N^{\mu },  \label{tv96b1}
\end{equation}%
which again is not satisfied since%
\begin{eqnarray}
&&\varepsilon _{\lambda \alpha \beta \gamma \delta }T^{\theta \lambda
}\left( \partial ^{\left[ \alpha \right. }t^{\left. \beta \gamma \delta %
\right] \vert \rho }\right) \sigma _{\theta \rho }=2\varepsilon _{\lambda
\alpha \beta \gamma \delta }\left[ \left( \partial ^{\lambda }t^{\alpha \xi
}\right) \left( \partial _{\xi }\partial _{\theta }t^{\beta \gamma \delta
\vert \theta }\right) \right.  \notag \\
&&\left. +\left( \partial ^{\lambda }t^{\alpha \beta }\right) \left(
\partial _{\xi }\partial _{\theta }t^{\xi \gamma \delta \vert \theta
}\right) \right] +\partial _{\mu }\left\{ \varepsilon _{\lambda \alpha \beta
\gamma \delta }\left[ \left( -\partial ^{\mu }t^{\lambda \alpha }+2\partial
^{\lambda }t^{\alpha \mu }\right. \right. \right.  \notag \\
&&\left. -\frac{1}{2}\partial _{\xi }t^{\lambda \alpha \mu \vert \xi
}\right) \partial _{\theta }t^{\beta \gamma \delta \vert \theta }-\frac{3}{2}%
\left( \left( \partial ^{\mu }t^{\lambda \alpha }\right) \partial ^{\beta
}t^{\gamma \delta }-t^{\lambda \alpha }\partial ^{\mu }\partial ^{\beta
}t^{\gamma \delta }+\right.  \notag \\
&&\left. \left. \left. \sigma ^{\beta \mu }t^{\lambda \alpha }\Box t^{\gamma
\delta }\right) -6\sigma ^{\beta \mu }t^{\lambda \alpha }\partial _{\theta
}\partial ^{\gamma }t^{\delta \theta }\right] \right\} ,  \label{tv96b2}
\end{eqnarray}%
and the first two terms in the right-hand side of (\ref{tv96b2}) do
not reduce to a full divergence. In conclusion, (\ref{tv96b}) takes
place only for $c_{3}=0$. Suppose now that
$a_{0c_{4}}^{\mathrm{t-A}}$ as solution to the Eq. (\ref{tv96c})
exists. The definition (\ref{tv49}) yields that it can be
represented as a sum of terms, each of them being quadratic in the
vector field, linear in $t_{\lambda \mu \nu \vert \alpha }$, and
containing
exactly two spatiotemporal derivatives. Moreover, each term may depend on $%
t_{\lambda \mu \nu \vert \alpha }$ only through gauge-invariant combinations
since otherwise $\gamma a_{0c_{4}}^{\mathrm{t-A}}$ would imply ghosts of
pure ghost number one from the $t$-sector, which is forbidden by the
expression of the left-hand side of (\ref{tv96c}). But the most general,
gauge-invariant quantities built out of $t_{\lambda \mu \nu \vert \alpha }$
with precisely two derivatives are proportional with the components of the
curvature tensor (\ref{tv20}). Consequently, we can write (up to
insignificant, full divergences) that%
\begin{equation}
a_{0c_{4}}^{\mathrm{t-A}}=c_{4}K^{\lambda \mu \nu \rho \vert \alpha \beta
}A_{\theta }A_{\xi }f_{\ \ \ \ \lambda \mu \nu \rho \alpha \beta }^{\theta
\xi },  \label{tv96c1}
\end{equation}%
where $f_{\ \ \ \ \lambda \mu \nu \rho \alpha \beta }^{\theta \xi }$ are
some non-derivative constants, symmetric in their upper indices. Acting with
$\gamma $ on (\ref{tv96c1}), replacing the resulting expression in (\ref%
{tv96c}), recalling that $f_{\ \ \ \ \lambda \mu \nu \rho \alpha \beta
}^{\theta \xi }$ cannot include Levi-Civita symbols (due to the Bianchi I
identities for the curvature tensor), and using the relations
\begin{equation}
T^{\lambda \mu }=\frac{\left( 4-D\right) }{2}K^{\lambda \mu \nu \rho \vert
\alpha \beta }\sigma _{\rho \beta }\sigma _{\nu \alpha },\qquad \partial
_{\lambda }T^{\lambda \mu }\equiv 0,  \label{reltk}
\end{equation}%
combined with the Bianchi II identities (\ref{bianchi2}), we obtain that (%
\ref{tv96c}) is satisfied if and only if $c_{4}=0$.

Based on the last results (setting $c_{2}=c_{3}=c_{4}=0$ in (\ref{tv91})--(%
\ref{tv95})) we can state that $a^{\mathrm{t-A}}$ actually stops at
antighost number one%
\begin{equation}
a^{\mathrm{t-A}}=a_{0}^{\mathrm{t-A}}+a_{1}^{\mathrm{t-A}},  \label{tv98}
\end{equation}%
with
\begin{equation}
a_{1}^{\mathrm{t-A}}=\delta _{D5}c_{5}\varepsilon _{\lambda \alpha \beta
\gamma \delta }A^{\ast \lambda }\mathcal{F}^{\alpha \beta \gamma \delta },
\label{tv99}
\end{equation}%
\begin{equation}
a_{0}^{\mathrm{t-A}}=-\frac{2c_{5}}{3}\delta _{D5}\varepsilon _{\mu \nu
\alpha \beta \gamma }F^{\mu \nu }\partial ^{\left[ \rho \right. }t^{\left.
\alpha \beta \gamma \right] \vert \theta }\sigma _{\theta \rho }+\bar{a}%
_{0}^{\mathrm{t-A}},  \label{tv100}
\end{equation}%
where $\bar{a}_{0}^{\mathrm{t-A}}$ is the general solution to the
`homogeneous' equation
\begin{equation}
\gamma \bar{a}_{0}^{\mathrm{t-A}}=\partial _{\mu }\overset{(0)}{\bar{m}}_{%
\mathrm{t-A}}^{\mu }.  \label{tv101}
\end{equation}%
We stress that here, at antighost number zero, we cannot replace the
equation $\gamma \bar{a}_{0}^{\mathrm{t-A}}=\partial _{\mu }\overset{(0)}{%
\bar{m}}_{\mathrm{t-A}}^{\mu }$ with the simpler one $\gamma \bar{a}_{0}^{%
\mathrm{t-A}}=0$ as we did before at strictly positive values of the
antighost number. For details, see Corollary 1 from the Appendix A in \cite%
{noijhep31}.

Next, we investigate the solutions to (\ref{tv101}). There are two main
types of solutions to this equation. The first type, to be denoted by $\bar{a%
}_{0}^{\prime \mathrm{t-A}}$, corresponds to $\overset{(0)}{\bar{m}}_{%
\mathrm{t-A}}^{\mu }=0$ and is given by gauge-invariant, non-integrated
densities constructed out of the original fields and their spatiotemporal
derivatives, which, according to (\ref{tv81}), are of the form $\bar{a}%
_{0}^{\prime \mathrm{t-A}}=\bar{a}_{0}^{\prime \mathrm{t-A}}\left( \left[
K_{\lambda \mu \nu \xi \vert \alpha \beta }\right] ,\left[ F_{\mu \nu }%
\right] \right) $, up to the condition that they effectively describe
cross-couplings between the two types of fields and cannot be written in a
divergence-like form. Such a solution implies at least three derivatives of
the fields and consequently must be forbidden by setting $\bar{a}%
_{0}^{\prime \mathrm{t-A}}=0$ since it breaks the hypothesis on the maximum
derivative order.

The second kind of solutions is associated with $\overset{(0)}{\bar{m}}_{%
\mathrm{t-A}}^{\mu }\neq 0$ in (\ref{tv101}), being understood that we
discard the divergence-like quantities and maintain the condition on the
maximum derivative order of the interacting Lagrangian being equal to two.
In order to solve this equation we start from the requirement that $\bar{a}%
_{0}^{\mathrm{t-A}}$ may contain at most two derivatives, so it can be
decomposed like
\begin{equation}
\bar{a}_{0}^{\mathrm{t-A}}=\omega _{0}+\omega _{1}+\omega _{2},
\label{tww60}
\end{equation}%
where $\left( \omega _{i}\right) _{i=\overline{0,2}}$ contains $i$
derivatives. Due to the different number of derivatives in the components $%
\omega _{0}$, $\omega _{1}$, and $\omega _{2}$, the Eq. (\ref{tww60}) is
equivalent to three independent equations
\begin{equation}
\gamma \omega _{k}=\partial _{\mu }j_{k}^{\mu },\quad k=0,1,2.  \label{twwxy}
\end{equation}

For $k=0$ the Eq. (\ref{twwxy}) implies the (necessary) conditions
\begin{equation}
\partial _{\lambda }\left( \frac{\partial \omega _{0}}{\partial t_{\lambda
\mu \nu \vert \alpha }}\right) =0,\qquad \partial _{\alpha }\left( \frac{%
\partial \omega _{0}}{\partial t_{\lambda \mu \nu \vert \alpha }}\right)
=0,\qquad \partial _{\mu }\left( \frac{\partial \omega _{0}}{\partial A_{\mu
}}\right) =0,  \label{tvcond0}
\end{equation}%
whose solutions read as
\begin{equation}
\frac{\partial \omega _{0}}{\partial t_{\lambda \mu \nu \vert \alpha }}%
=0,\qquad \frac{\partial \omega _{0}}{\partial A_{\mu }}=0,  \label{tvsol0}
\end{equation}%
so $\omega _{0}$ provides no cross-couplings between $t_{\lambda \mu \nu
\vert \alpha }$ and $A_{\mu }$, and therefore we can take
\begin{equation}
\omega _{0}=0  \label{omega0}
\end{equation}%
in (\ref{tww60}). The solution to the more general equations (\ref{tv32c})
is of the form (\ref{tv32d}), but it does not apply to the first two
equations in (\ref{tvcond0}) as $\partial \omega _{0}/\partial t_{\lambda
\mu \nu \vert \alpha }$ are by hypothesis derivative-free. The same
observation is valid with respect to the equation $\partial _{\mu }M^{\mu
}=0 $, whose solution reads as $M^{\mu }=\partial _{\nu }N^{\nu \mu }$, with
$N^{\nu \mu } $ antisymmetric, so it cannot enter the second solution from (%
\ref{tvsol0}) since $\partial \omega _{0}/\partial A_{\mu }$ has no
derivatives.

For $k=1$ the Eq. (\ref{twwxy}) leads to the requirements
\begin{equation}
\partial _{\lambda }\left( \frac{\delta \omega _{1}}{\delta t_{\lambda \mu
\nu \vert \alpha }}\right) =0,\qquad \partial _{\alpha }\left( \frac{\delta
\omega _{1}}{\delta t_{\lambda \mu \nu \vert \alpha }}\right) =0,\qquad
\partial _{\mu }\left( \frac{\delta \omega _{1}}{\delta A_{\mu }}\right) =0,
\label{tvcond1}
\end{equation}%
where $\delta \omega _{1}/\delta t_{\lambda \mu \nu \vert \alpha }$ denote
the Euler-Lagrange derivatives of $\omega _{1}$. Because $\omega _{1}$ is by
hypothesis of order one in the spatiotemporal derivatives of the fields, the
arguments presented in relation with the case $k=0$ provide the solutions
\begin{equation}
\frac{\delta \omega _{1}}{\delta t_{\lambda \mu \nu \vert \alpha }}=0,\quad
\frac{\delta \omega _{1}}{\delta A_{\mu }}=\partial _{\nu }B^{\nu \mu },
\label{tvsol1}
\end{equation}%
where the antisymmetric functions $B^{\nu \mu }$ have no derivatives. The
first solution forbids the cross-couplings between the two types of fields,
allowing only the self-interactions of the vector field with precisely one
derivative, so we can safely take
\begin{equation}
\omega _{1}=0.  \label{omega1}
\end{equation}

We pass now to the Eq. (\ref{twwxy}) for $k=2$, which produces the
restrictions
\begin{equation}
\partial _{\lambda }\left( \frac{\delta \omega _{2}}{\delta t_{\lambda \mu
\nu \vert \alpha }}\right) =0,\qquad \partial _{\alpha }\left( \frac{\delta
\omega _{2}}{\delta t_{\lambda \mu \nu \vert \alpha }}\right) =0,\qquad
\partial _{\mu }\left( \frac{\delta \omega _{2}}{\delta A_{\mu }}\right) =0,
\label{tvcond2}
\end{equation}%
whose solution, by virtue of the discussion made at the case $k=0$, is%
\begin{equation}
\frac{\delta \omega _{2}}{\delta t_{\lambda \mu \nu \vert \alpha }}=\partial
_{\rho }\partial _{\beta }U^{\lambda \mu \nu \rho \vert \alpha \beta
},\qquad \frac{\delta \omega _{2}}{\delta A_{\mu }}=\partial _{\nu }\Phi
^{\nu \mu },  \label{tvsol2}
\end{equation}%
where $U^{\lambda \mu \nu \rho \vert \alpha \beta }$ has the mixed symmetry
of the curvature tensor (\ref{tv20}) and has no derivatives, while $\Phi
^{\nu \mu }$ is antisymmetric and comprises one spatiotemporal derivative of
the fields. At this stage it is useful to introduce a derivation in the
algebra of the fields and of their derivatives, which counts the powers of
the fields and their derivatives
\begin{eqnarray}
N &=&\sum\limits_{k\geq 0}\left( \left( \partial _{\mu _{1}}\cdots \partial
_{\mu _{k}}t_{\lambda \mu \nu \vert \alpha }\right) \frac{\partial }{%
\partial \left( \partial _{\mu _{1}}\cdots \partial _{\mu _{k}}t_{\lambda
\mu \nu \vert \alpha }\right) }\right.  \notag \\
&&\left. +\left( \partial _{\mu _{1}}\cdots \partial _{\mu _{k}}A_{\mu
}\right) \frac{\partial }{\partial \left( \partial _{\mu _{1}}\cdots
\partial _{\mu _{k}}A_{\mu }\right) }\right) ,  \label{tww74}
\end{eqnarray}%
so for every non-integrated density $\chi $ we have that
\begin{equation}
N\chi =t_{\lambda \mu \nu \vert \alpha }\frac{\delta \chi }{\delta
t_{\lambda \mu \nu \vert \alpha }}+A_{\mu }\frac{\delta \chi }{\delta A_{\mu
}}+\partial _{\mu }s^{\mu }.  \label{tww75}
\end{equation}%
If $\chi ^{\left( l\right) }$ is a homogeneous polynomial of order $l>0$ in
the fields $\left\{ t_{\lambda \mu \nu \vert \alpha },A_{\mu }\right\} $ and
their derivatives, then $N\chi ^{\left( l\right) }=l\chi ^{\left( l\right) }$%
. Using (\ref{tvsol2}) and (\ref{tww75}), we find that
\begin{equation}
N\omega _{2}=\frac{1}{8}K_{\lambda \mu \nu \rho \vert \alpha \beta
}U^{\lambda \mu \nu \rho \vert \alpha \beta }-\frac{1}{2}F_{\mu \nu }\Phi
^{\mu \nu }+\partial _{\mu }v^{\mu }.  \label{tww76a}
\end{equation}%
We expand $\omega _{2}$ according to the various eigenvalues of $N$ like%
\begin{equation}
\omega _{2}=\sum\limits_{l>0}\omega _{2}^{\left( l\right) },  \label{tww77}
\end{equation}%
where $N\omega _{2}^{\left( l\right) }=l\omega _{2}^{\left( l\right) }$,
such that
\begin{equation}
N\omega _{2}=\sum\limits_{l>0}l\omega _{2}^{\left( l\right) }.  \label{tww78}
\end{equation}%
Comparing (\ref{tww76a}) with (\ref{tww78}), we reach the conclusion that
the decomposition (\ref{tww77}) induces a similar decomposition with respect
to $U^{\lambda \mu \nu \rho \vert \alpha \beta }$ and $\Phi ^{\mu \nu }$
\begin{equation}
U^{\lambda \mu \nu \rho \vert \alpha \beta }=\sum\limits_{l>0}U_{\left(
l-1\right) }^{\lambda \mu \nu \rho \vert \alpha \beta },\;\Phi ^{\mu \nu
}=\sum\limits_{l>0}\Phi _{\left( l-1\right) }^{\mu \nu }.  \label{tww79}
\end{equation}%
Substituting (\ref{tww79}) into (\ref{tww76a}) and comparing the resulting
expression with (\ref{tww78}), we obtain that
\begin{equation}
\omega _{2}^{\left( l\right) }=\frac{1}{8l}K_{\lambda \mu \nu \rho \vert
\alpha \beta }U_{\left( l-1\right) }^{\lambda \mu \nu \rho \vert \alpha
\beta }-\frac{1}{2l}F_{\mu \nu }\Phi _{\left( l-1\right) }^{\mu \nu
}+\partial _{\mu }\bar{v}_{(l)}^{\mu }.  \label{tvprform}
\end{equation}%
Introducing (\ref{tvprform}) in (\ref{tww77}), we arrive at
\begin{equation}
\omega _{2}=K_{\lambda \mu \nu \rho \vert \alpha \beta }\bar{U}^{\lambda \mu
\nu \rho \vert \alpha \beta }+F_{\mu \nu }\bar{\Phi}^{\mu \nu }+\partial
_{\mu }\bar{v}^{\mu },  \label{tww81}
\end{equation}%
where
\begin{equation}
\bar{U}^{\lambda \mu \nu \rho \vert \alpha \beta }=\sum\limits_{l>0}\frac{1}{%
8l}U_{\left( l-1\right) }^{\lambda \mu \nu \rho \vert \alpha \beta },\;\bar{%
\Phi}^{\mu \nu }=-\sum\limits_{l>0}\frac{1}{2l}\Phi _{\left( l-1\right)
}^{\mu \nu }.  \label{tww82}
\end{equation}%
Applying $\gamma $ on (\ref{tww81}), after long and tedious computations we
infer that a necessary condition for the existence of solutions to the
equation $\gamma \omega _{2}=\partial _{\mu }j_{2}^{\mu }$ is that the
functions $\bar{U}^{\lambda \mu \nu \rho \vert \alpha \beta }$ and $\bar{\Phi%
}^{\mu \nu }$ entering (\ref{tww81}) have the expressions
\begin{equation}
\bar{U}^{\lambda \mu \nu \rho \vert \alpha \beta }=C^{\lambda \mu \nu \rho
;\alpha \beta ;\sigma }A_{\sigma },\;\bar{\Phi}^{\mu \nu }=\bar{k}^{\mu \nu
\rho ;\alpha \beta \gamma ;\lambda }\partial _{\rho }t_{\alpha \beta \gamma
\vert \lambda },  \label{tvsol2a}
\end{equation}%
where $C^{\lambda \mu \nu \rho ;\alpha \beta ;\sigma }$ and $\bar{k}^{\mu
\nu \rho ;\alpha \beta \gamma ;\lambda }$ are non-derivative constants,
antisymmetric in the indices followed by or between semicolons. Substituting
(\ref{tvsol2a}) in (\ref{tww81}) we deduce
\begin{equation}
\omega _{2}=C^{\lambda \mu \nu \rho ;\alpha \beta ;\sigma }K_{\lambda \mu
\nu \rho \vert \alpha \beta }A_{\sigma }+\partial _{\rho }\left( F_{\mu \nu }%
\bar{k}^{\mu \nu \rho ;\alpha \beta \gamma ;\lambda }t_{\alpha \beta \gamma
\vert \lambda }+\bar{v}^{\rho }\right) .  \label{tvsol2b}
\end{equation}%
Applying once more $\gamma $ on (\ref{tvsol2b}), we find that the equation $%
\gamma \omega _{2}=\partial _{\mu }j_{2}^{\mu }$ holds if and only if
\begin{equation}
C^{\lambda \mu \nu \rho ;\alpha \beta ;\sigma }\partial _{\sigma }K_{\lambda
\mu \nu \rho \vert \alpha \beta }=0.  \label{condC}
\end{equation}%
Taking into account the fact that the only vanishing combinations
constructed from the first-order derivatives of the curvature tensor are the
Bianchi II identities (\ref{bianchi2}) and their traces, we find that the
constants $C^{\lambda \mu \nu \rho ;\alpha \beta ;\sigma }$ must
simultaneously ensure (\ref{condC}) and also a non-vanishing term in (\ref%
{tvsol2b}). In $D\geq 5$ dimensions there are no such constants, so we must
take $C^{\lambda \mu \nu \rho ;\alpha \beta ;\sigma }=0$. Eliminating the
(trivial) divergence from (\ref{tvsol2b}), we can state that
\begin{equation}
\omega _{2}=0.  \label{omega2}
\end{equation}%
Replacing (\ref{omega0}), (\ref{omega1}), and (\ref{omega2}) in (\ref{tww60}%
), we finally have that
\begin{equation}
\bar{a}_{0}^{\mathrm{t-A}}=0  \label{tv102}
\end{equation}%
in (\ref{tv100}).

Inserting now the results (\ref{tv99})--(\ref{tv100}) and (\ref{tv102}) in
the relation (\ref{tv98}) and the resulting expression together with (\ref%
{tv86a}) and (\ref{tv86b}) into the formula (\ref{tv84}), we obtain that the
most general form of the first-order deformation associated with the free
theory (\ref{tv1}) is given by%
\begin{eqnarray}
S_{1} &=&\int d^{5}x\,\varepsilon ^{\lambda \mu \nu \rho \kappa }\left[
c_{5}\left( A_{\lambda }^{\ast }\mathcal{F}_{\mu \nu \rho \kappa }-\frac{2}{3%
}F_{\lambda \mu }\partial _{\left[ \xi \right. }t_{\left. \nu \rho \kappa %
\right] \vert \theta }\sigma ^{\theta \xi }\right) \right.  \notag \\
&&\left. +\frac{c}{3\cdot 4!}F_{\lambda \mu }F_{\nu \rho }A_{\kappa }\right]
,  \label{tv103}
\end{eqnarray}%
being parametrized by just two real (and so far arbitrary) constants.

\section{Higher-order deformations\label{b}}

In the sequel we approach the higher-order deformation equations. The
second-order deformation is controlled by the Eq. (\ref{tv65}), whose
solution, on behalf of the result (\ref{tv103}), is expressed by
\begin{eqnarray}
S_{2} &=&\int d^{5}x\left[ \frac{16}{3}c_{5}^{2}\left( \partial
_{\left[ \xi \right. }t_{\left. \nu \rho \kappa \right] \vert \theta
}\sigma ^{\theta \xi
}\right) \partial ^{\left[ \xi ^{\prime }\right. }t^{\left. \nu \rho \kappa %
\right] \vert \theta ^{\prime }}\sigma _{\theta ^{\prime }\xi
^{\prime }}\right.
\notag \\
&&\left. -\frac{2}{3}cc_{5}F^{\lambda \mu }\left( A^{\rho }\partial
_{\left[ \xi \right. }t_{\left. \lambda \mu \rho \right] \vert
\theta }\sigma ^{\theta \xi }-t_{\lambda \mu }^{\ast }\eta \right)
\right] .  \label{tv104}
\end{eqnarray}%
Using (\ref{tv103})--(\ref{tv104}) in (\ref{tv66}) we infer the third-order
deformation as%
\begin{eqnarray}
S_{3} &=&\frac{4}{3}cc_{5}^{2}\int d^{5}x\varepsilon ^{\lambda \mu
\nu \rho \kappa }\left[ \left( \frac{1}{3}\mathcal{G}_{\lambda
\alpha \vert \beta }^{\ast }\sigma ^{\alpha \beta }\eta +t_{\alpha
\lambda }^{\ast }A^{\alpha }\right)
\mathcal{F}_{\mu \nu \rho \kappa }\right.  \notag \\
&&\left. +\frac{2}{3}\left( 2t_{\lambda \mu }^{\ast }\eta -A^{\alpha
}\partial _{\left[ \xi ^{\prime }\right. }t_{\left. \alpha \lambda \mu %
\right] \vert \theta ^{\prime }}\sigma ^{\theta ^{\prime }\xi
^{\prime }}\right)
\partial _{\left[ \xi \right. }t_{\left. \nu \rho \kappa \right] \vert \theta
}\sigma ^{\theta \xi }\right] .  \label{tv105}
\end{eqnarray}%
Substituting the expressions (\ref{tv103})--(\ref{tv105}) into the Eq. (\ref%
{tv66a}), we obtain the equivalent relation%
\begin{eqnarray}
&&64cc_{5}^{3}\int d^{5}x\left( 2t_{\lambda \mu }^{\ast }\mathcal{F}%
^{\lambda \alpha \beta \gamma }\mathcal{F}_{\quad \alpha \beta
\gamma }^{\mu }-\partial _{\left[ \xi ^{\prime }\right. }t_{\left.
\alpha \beta \mu \right] \vert \theta ^{\prime }}\sigma ^{\theta
^{\prime }\xi ^{\prime }}\partial _{\left[ \xi \right. }t_{\left.
\gamma \delta \nu \right] \vert \theta }\sigma ^{\theta \xi }\sigma
^{\mu \nu }\mathcal{F}^{\alpha \beta \gamma \delta }\right)
\notag \\
&&+s\left[ \frac{4}{9}cc_{5}^{2}\int d^{5}x\left( \frac{c}{3}A^{\left[
\lambda \right. }F^{\left. \mu \nu \right] }A_{\lambda }F_{\mu \nu
}-32c_{5}t_{\lambda \mu }^{\ast }F^{\lambda \mu }\eta \right) +S_{4}\right]
=0.  \label{tv106}
\end{eqnarray}%
If we make the notations $S_{4}=\int d^{5}x\,b$,
\begin{equation}
64cc_{5}^{3}\left( 2t_{\lambda \mu }^{\ast }\mathcal{F}^{\lambda \alpha
\beta \gamma }\mathcal{F}_{\quad \alpha \beta \gamma }^{\mu }-\partial _{%
\left[ \xi ^{\prime }\right. }t_{\left. \alpha \beta \mu \right]
\vert \theta ^{\prime }}\sigma ^{\theta ^{\prime }\xi ^{\prime
}}\partial _{\left[ \xi \right. }t_{\left. \gamma \delta \nu \right]
\vert \theta }\sigma ^{\theta \xi }\sigma ^{\mu \nu
}\mathcal{F}^{\alpha \beta \gamma \delta }\right) \equiv \Lambda ,
\label{tvnotlambda}
\end{equation}%
and%
\begin{equation}
\frac{4}{9}cc_{5}^{2}\left( \frac{c}{3}A^{\left[ \lambda \right. }F^{\left.
\mu \nu \right] }A_{\lambda }F_{\mu \nu }-32c_{5}t_{\lambda \mu }^{\ast
}F^{\lambda \mu }\eta \right) +b\equiv \bar{b},  \label{tvnotb}
\end{equation}%
then (\ref{tv106}) takes the local form%
\begin{equation}
\Lambda +\partial _{\mu }w^{\mu }+s\bar{b}=0,  \label{tv106a}
\end{equation}%
with
\begin{equation}
\mathrm{gh}\left( \Lambda \right) =1,\qquad \mathrm{gh}\left( \bar{b}\right)
=0,\qquad \mathrm{gh}\left( w^{\mu }\right) =1.  \label{tv106b}
\end{equation}%
From (\ref{tvnotlambda}) we see that $\Lambda $ decomposes like
\begin{equation}
\Lambda =\Lambda _{0}+\Lambda _{1},\qquad \mathrm{agh}\left( \Lambda
_{i}\right) =i,\qquad i=0,1,  \label{tv106c}
\end{equation}%
with
\begin{eqnarray}
\Lambda _{0} &=&-64cc_{5}^{3}\partial _{\left[ \xi ^{\prime }\right.
}t_{\left. \alpha \beta \mu \right] \vert \theta ^{\prime }}\sigma
^{\theta ^{\prime }\xi ^{\prime }}\partial _{\left[ \xi \right.
}t_{\left. \gamma
\delta \nu \right] \vert \theta }\sigma ^{\theta \xi }\sigma ^{\mu \nu }\mathcal{F%
}^{\alpha \beta \gamma \delta },  \label{tv107a} \\
\Lambda _{1} &=&128cc_{5}^{3}t_{\lambda \mu }^{\ast }\mathcal{F}^{\lambda
\alpha \beta \gamma }\mathcal{F}_{\quad \alpha \beta \gamma }^{\mu }.
\label{tv107b}
\end{eqnarray}%
Employing the decomposition (\ref{tv39}) of the BRST differential and (\ref%
{tv106c}), it results that in (\ref{tv106a}) we can take (without loss of
generality) $\bar{b}$ and $w^{\mu }$ to stop at antighost number two%
\begin{eqnarray}
\bar{b} &=&\bar{b}_{0}+\bar{b}_{1}+\bar{b}_{2},\qquad \mathrm{agh}\left(
\bar{b}_{i}\right) =i,\qquad i=0,1,2,  \label{tv108a} \\
w^{\mu } &=&w_{0}^{\mu }+w_{1}^{\mu }+w_{2}^{\mu },\qquad \mathrm{agh}\left(
w_{i}^{\mu }\right) =i,\qquad i=0,1,2.  \label{tv108b}
\end{eqnarray}%
By projecting the Eq. (\ref{tv106a}) on the various values of the antighost
number, we infer an equivalent tower of equations%
\begin{eqnarray}
0 &=&\gamma \bar{b}_{2}+\partial _{\mu }w_{2}^{\mu },  \label{tv109a} \\
\Lambda _{1} &=&-\left( \delta \bar{b}_{2}+\gamma \bar{b}_{1}\right)
-\partial _{\mu }w_{1}^{\mu },  \label{tv109b} \\
\Lambda _{0} &=&-\left( \delta \bar{b}_{1}+\gamma \bar{b}_{0}\right)
-\partial _{\mu }w_{0}^{\mu }.  \label{tv109c}
\end{eqnarray}%
According to the general result from Corollary 1, Appendix A in \cite%
{noijhep31}, we can always replace (\ref{tv109a}) with the simpler equation%
\begin{equation}
\gamma \bar{b}_{2}=0  \label{tv110}
\end{equation}%
and take the current $w^{\mu }$ from (\ref{tv108b}) to stop at
antighost number one, $w_{2}^{\mu }=0$. Looking at (\ref{tv107b}),
we can state that both $\bar{b}_{2}$ and $\bar{b}_{1}$ must contain
only BRST generators from the $(3,1)$ sector. The solution to
(\ref{tv110}) results from (\ref{hgammat}) at antighost number
$I=2$. However, there is no a priori reason to consider the
corresponding invariant polynomial entering $\bar{b}_{2}$ to pertain
to $H_{2}\left( \delta \vert d\right) $. On the other hand,
(\ref{tv107b}) emphasizes that in order to render $\bar{b}_{2}$ to
contribute nontrivially to $\Lambda _{1}$ through the Eq.
(\ref{tv109b}), it is necessary that the above mentioned invariant
polynomial contains \emph{only} the antifields $\mathcal{G}^{\prime
\ast \lambda \mu \vert \vert \nu }\sigma _{\mu \nu }$ (and neither
products of two antifields $t^{\ast \lambda \mu \nu \vert \alpha } $
nor the curvature tensor $K_{\lambda \mu \nu \xi \vert \alpha \beta
}$). Due to the fact that $\mathcal{G}^{\prime \ast \lambda \mu
\vert \vert \nu }\sigma _{\mu \nu } $ belongs to
$H_{2}^{\mathrm{inv}}\left( \delta \vert d\right) $, it follows that
we have reduced the problem of solving the Eq. (\ref{tv110}) to the
corresponding problem from the case of constructing
self-interactions for the tensor $t_{\lambda \mu \nu \vert \alpha
}$. As it has been shown in \cite{noijhep31} under the same
hypotheses like those employed here, the solution to (\ref{tv110})
reads
\begin{equation}
\bar{b}_{2}=0,  \label{b2}
\end{equation}%
so the Eq. (\ref{tv109b}) takes the form%
\begin{equation}
\Lambda _{1}=-\gamma \bar{b}_{1}-\partial _{\mu }w_{1}^{\mu }.  \label{tv111}
\end{equation}%
The formula (\ref{tv107b}) emphasizes that $\Lambda _{1}$ is a
nontrivial co-cycle from $H\left( \gamma \right) $, which does not
reduce to a $\gamma $-exact modulo $d$ term, such that we must take
its coefficient to vanish
\begin{equation}
cc_{5}^{3}=0.  \label{tv114}
\end{equation}%
So far, we have shown that the existence of $S_{4}$ as solution to the Eq. (%
\ref{tv106}) requires the condition (\ref{tv114}).

There appear two main cases related to the solutions of
(\ref{tv114}). If we take
\begin{equation}
c_{5}=0  \label{tv115}
\end{equation}%
and leave $c$ to be an arbitrary, real constant (for definiteness,
we fix this constant to unity, $c=1$), then from (\ref{tv106}) it
follows that we can set
\begin{equation}
S_{4}=0.  \label{s4}
\end{equation}%
In the meantime, from (\ref{tv104})--(\ref{tv105}) we get that
\begin{equation}
S_{2}=0=S_{3}.  \label{s2s3}
\end{equation}%
The results (\ref{s4}) and (\ref{s2s3}) ensure that we can actually
take all the other higher-order deformations to vanish,
\begin{equation}
S_{k}=0,\qquad k>4.  \label{sk4}
\end{equation}%
In this situation (\ref{tv103}) and (\ref{tv115}) produce the first-order
deformation like%
\begin{equation}
S_{1}=\frac{1}{3\cdot 4!}\int d^{5}x\,\varepsilon ^{\lambda \mu \nu \rho
\kappa }F_{\lambda \mu }F_{\nu \rho }A_{\kappa }.  \label{tv116}
\end{equation}%
The second possibility is to work with%
\begin{equation}
c=0  \label{tv118}
\end{equation}%
and take $c_{5}$ to be an arbitrary real constant (for definiteness, we fix
this constant to unity, $c_{5}=1$). Consequently, from (\ref{tv105})--(\ref%
{tv106}) we find that
\begin{equation}
S_{k}=0,\qquad k\geq 3.  \label{sk}
\end{equation}%
The first- and second-order deformations result from (\ref{tv103}) and (\ref%
{tv104}) where we set (\ref{tv118}) and $c_{5}=1$, and take the form%
\begin{eqnarray}
S_{1} &=&\int d^{5}x\,\varepsilon ^{\lambda \mu \nu \rho \kappa }\left(
A_{\lambda }^{\ast }\mathcal{F}_{\mu \nu \rho \kappa }-\frac{2}{3}F_{\lambda
\mu }\partial _{\left[ \xi \right. }t_{\left. \nu \rho \kappa \right]
\vert \theta }\sigma ^{\theta \xi }\right) ,  \label{s1c0} \\
S_{2} &=&\frac{16}{3}\int d^{5}x\left( \partial _{\left[ \xi \right.
}t_{\left. \nu \rho \kappa \right] \vert \theta }\sigma ^{\theta \xi
}\right)
\partial ^{\left[ \xi ^{\prime }\right. }t^{\left. \nu \rho \kappa \right]
\vert \theta ^{\prime }}\sigma _{\theta ^{\prime }\xi ^{\prime }}.
\label{s2c0}
\end{eqnarray}%
Formulas (\ref{s4})--(\ref{tv116}) lead to the full deformed solution of the
classical master equation as in (\ref{tv117}). Similarly, relations (\ref{sk}%
)--(\ref{s2c0}) yield the overall deformed solution of the form (\ref{tv119}%
).

\end{document}